\documentclass[preprint,3p,11pt,sort&compress]{elsarticle}
\usepackage[T1]{fontenc}
\usepackage[utf8]{inputenc}
\usepackage{amsmath}
\usepackage{amssymb, bm}
\usepackage{hyperref}
\usepackage{xcolor}
\usepackage{upgreek}
\usepackage[linesnumbered,ruled,vlined]{algorithm2e}
\usepackage{textcomp}
\usepackage{setspace}
\usepackage{lineno}
\usepackage{booktabs}
\usepackage{threeparttable}
\usepackage{array}
\usepackage{dsfont}
\newcommand{\ind}{\mathds{1}}
\usepackage{longtable}
\usepackage{endnotes}

\journal{arXiv}

\begin{document}

\begin{frontmatter}

\title{Value of Information-based assessment of strain-based thickness loss monitoring in ship hull structures}

\author[NTUA]{Nicholas E. Silionis\texorpdfstring{\corref{cor}}{}}
\ead{nsil@naval.ntua.gr}
\author[NTUA]{Konstantinos N. Anyfantis}
\ead{kanyf@naval.ntua.gr}

\affiliation[NTUA]{organization={Ship-Hull Structural Health Monitoring (S-H SHM) Group, School of Naval Architecture and Marine Engineering, National Technical University of Athens},
            addressline={9 Heroon Polytechniou Av.}, 
            city={Athens},
            postcode={15780 Zografos}, 
            country={Greece}}

\cortext[cor]{Corresponding author.}

\begin{abstract}
Recent advances in Structural Health Monitoring (SHM) have attracted industry interest, yet real-world applications, such as in ship structures, remain scarce. Despite SHM’s potential to optimise maintenance, its adoption in ships is limited due to the lack of clearly quantifiable benefits for hull structural maintenance. This study employs a Bayesian pre-posterior decision analysis to quantify the value of information from SHM systems monitoring corrosion-induced thickness loss in ship hulls, in a first-of-its kind analysis for ship structures. We define decision-making consequence cost functions based on exceedance probabilities relative to a target thickness loss threshold, which can be set by the decision-maker. This introduces a practical aspect to our framework, that enables implicitly modelling the decision-maker’s risk perception. We apply this framework to a large-scale, high-fidelity numerical model of a commercial vessel and examine the relative benefits of different thickness loss monitoring strategies, including strain-based SHM and traditional on-site inspections.
\end{abstract}


\end{frontmatter}

\section{Introduction}\label{seq1}
Over the past decades, significant advances have been made in Structural Health Monitoring (SHM) research by capitalising on the simultaneous developments in sensing technology and the increase in processing power. Although the potential of SHM has been recognised by stakeholders across various engineering disciplines, industrial application at scale is still at a nascent stage. The shipping industry is a typical example; in recent years key stakeholders—primarily classification societies \citep{ABS1,ABS2,Smogeli2017}—have viewed the potential offered by SHM to optimise maintenance planning as highly promising and thus included it in their strategic goals.

Despite this interest, real-world installations of SHM systems are still quite rare and are typically found in non-civilian vessels \citep{Torkildsen2005}. Even then, the scope of these systems is confined to monitoring stresses on the ship hull and providing alarms if certain threshold values have been exceeded. Using structural response data to aid in decision-making or to optimise vessel operation and maintenance (O\&M) remains largely a research topic \citep{SilionisSome}.

Current hull structural maintenance practices centre on inspections conducted at fixed intervals, conventionally referred to as hull surveys. These occur every 1, 3, and 5 years, as mandated by regulations issued by classification societies. The timing of these inspections, and the actions undertaken during each, are precisely defined by class rules. For bulk carriers and oil tankers, these rules conform to unified standards set by the International Association of Classification Societies (IACS) \citep{IACS2021, IACS2007}. During these surveys, areas of the hull structure prone to specific deterioration mechanisms, such as corrosion or fatigue cracking, are inspected visually or through non-destructive testing methods, e.g., ultrasonic thickness measurement.

Locations that are largely inaccessible during operation, such as water ballast tanks in the vessel’s double bottom, are inspected infrequently—typically every five years during dry-docking. However, structural components in these areas are prone to significant deterioration, particularly corrosion, due to the harsh operational environment characterised by high salinity and temperature fluctuations.
The combination of severe corrosion potential, limited accessibility, and infrequent inspection has spurred research into condition-based maintenance, where knowledge of structural deterioration is used to schedule maintenance actions flexibly according to the predicted state of the structure at a given time.

Earlier works in this direction mainly applied statistical methods to describe corrosion-induced thickness loss using historical data collected from operating vessels. These methods used classic regression techniques to fit models for the temporal evolution of thickness loss, such as the modified linear model proposed by Paik et al. \citep{Paik2003}, or the exponential-type models proposed by Garbatov et al. \citep{Garbatov2010} and Qin et al. \citep{Qin2003}.

More recently, Bayesian inference has been used to explicitly incorporate measurement uncertainty into the inference process, enabling more robust estimation of thickness loss from limited inspection data \citep{Woloszyk2024}. Bayesian methods have also been used to obtain probabilistic variants of thickness loss models \citep{Georgiadis2019,Kim2022}, which enable uncertainty quantification, and are thus better suited for decision support, compared to deterministic point predictors.
Models fitted to historical data can be ill-suited for decision support, particularly for newly built structures. Variations in operational conditions or structural design, among other factors, may render models that effectively describe deterioration in older structures unsuitable for newer ones. Successful knowledge transfer between structural systems requires certain conditions related to both data collection and model selection. This remains an active research area closely linked to population-based SHM \citep{PBSHM1, PBSHM2, PBSHM3}. For corrosion-induced thickness loss in ship hulls, the quantity and quality of available data can pose a major challenge to population-based methods, as reflected by the significant measurement uncertainty reported in \citep{Paik2003,Woloszyk2024}.

This limitation, combined with the restricted accessibility of enclosed spaces for data collection, has led researchers to explore indirect monitoring of corrosion-induced thickness loss. This is a typical SHM task achieved using structural response data, such as strains measured at locations of interest. The authors have previously employed static strain data to detect thickness loss in simplified structural components \citep{Katsoudas2023} and realistic hull structures \citep{SilionisMS} subjected to stochastic loading. Ghasemzadeh et al. \citep{Ghasemzadeh2023_v1,Ghasemzadeh2023_v2} used the inverse finite element method (iFEM)—a strain-based approach—to identify pitting corrosion in lab-scale steel components used in marine applications.

These works demonstrate that SHM is a feasible approach for remote, indirect thickness loss detection—and potentially prediction. However, they do not fully establish it as a viable alternative for hull structural maintenance planning against this form of deterioration, as they provide no insight into the benefits of adopting SHM over traditional maintenance strategies. This question is central not only to marine structures but to SHM more broadly, and researchers have sought to address it through Bayesian decision theory, particularly using the concept of the Value of Information (VoI) \citep{Pozzi2011, Zonta2014, Thons2018, Straub_VoI, Straub2014,Thons2025,Kamariotis2024}.

The VoI provides a rational basis for estimating the expected capital gains from implementing an information acquisition strategy—such as conventional inspections or continuous monitoring—on an existing structure. In simple terms, it quantifies the cost savings achieved by using monitoring information to support operation and maintenance decisions. If these savings exceed the intrinsic cost of information acquisition and any associated maintenance or related expenses, the strategy can be considered beneficial.

As a result, the VoI has been widely applied in SHM across various engineering domains, with most studies focusing on civil infrastructure. Zonta et al. \citep{Zonta2014}—in one of the earliest works on the topic— demonstrated its use in evaluating the impact of monitoring on the management of a pedestrian bridge subject to potentially critical damage. Kamariotis et al. \citep{Kamariotis2022,Kamariotis2023} and Giordano et al. \citep{Giordano2020} used the VoI to assess the influence of monitoring information on the management of scoured bridges.

Iannacone et al. \citep{Iannacone2022} proposed a unified framework for quantifying the VoI from inspections and SHM procedures for a bridge subject to both gradual deterioration and shock events, while Song et al. \citep{Song2022} integrated partially observable Markov decision processes within the VoI formulation to address non-stationary deterioration. Chadha et al. \citep{Chadha2021,Chadha2023,Chadha2025} applied the VoI to case studies involving navigation lock miter gates, including recent work on optimal sensor placement (OSP) for cost-effective SHM.

The VoI has also been incorporated into OSP frameworks across diverse applications, including ultrasonic guided-wave SHM of aeronautical structures \citep{Cantero2019} and geotechnical structures for computationally efficient reliability updating \citep{Tian2025}. Recent studies have combined the VoI and Bayesian decision theory principles to address practical SHM challenges, such as time-varying system performance \citep{Zhang2023,Wu2024}, sensor reliability in system design \citep{Yang2023_MSSP}, and maintenance planning under model uncertainty \citep{Arcieri2023}.

Beyond civil infrastructure, Nielsen et al. \citep{Nielsen2021} used the VoI on a case study related to risk-based maintenance of an operating wind turbine blade. For marine and offshore structures, Zou et al. \citep{Zou2022} quantified the VoI from inspection outcomes in fatigue maintenance planning for marine structures, while Rezende et al. \citep{Rezende2024} developed reliability-based inspection planning methods for offshore mooring chains accounting for both fatigue and corrosion deterioration.

\begin{figure}[!htp]
\centering
\includegraphics[scale=1.0]{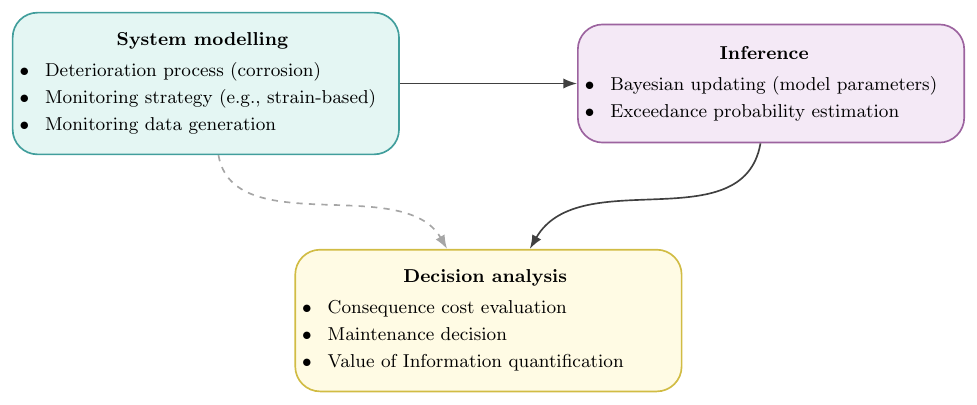}
\caption{Conceptual overview of the proposed VoI-based assessment framework for strain-based monitoring in ship hull structural maintenance planning.}
\label{high_level_flow}
\end{figure}

Although the VoI has been applied across various structural domains, its use for ship hull structures remains limited. To the authors’ knowledge, this study presents the first comprehensive VoI-based assessment of strain-based monitoring for ship hull structural maintenance planning. A high-level overview of the proposed VoI-based framework is shown in Figure~\ref{high_level_flow}. The objective is to demonstrate a methodology that enables stakeholders to translate their interest in SHM into actionable investment decisions, thereby advancing its practical implementation for ship hulls.

A Bayesian pre-posterior decision analysis is employed to quantify the VoI from various strain-based strategies for monitoring corrosion-induced thickness loss—a typical gradual deterioration process in ship hulls. A major challenge in this domain is the scarcity of operational monitoring data, which are often proprietary and rarely available publicly. This motivates the use of pre-posterior VoI frameworks that inform investment decisions using physics-based models to generate synthetic data prior to deployment. Here, a high-fidelity FE model of a commercial vessel developed by the authors \citep{SilionisMS} is used to produce the synthetic strain data required for the analysis. The VoI is quantified using the expected reward-to-investment risk ratio proposed by Chadha et al.~\citep{Chadha2021}.

The key methodological contribution is a modified consequence cost formulation for the binary decision setting. Unlike previous approaches that condition costs on state variables \citep{Chadha2021,Chadha2023}, costs are here defined directly using exceedance probabilities relative to a decision-maker-specified deterioration threshold. This formulation naturally links to practical maintenance planning while implicitly capturing risk perception through threshold selection. Consequence costs are expressed in normalised form (0–1) due to limited cost data availability, with extreme values determined through rational considerations.

The VoI is evaluated for different SHM strategies and decision settings, and compared with a scenario involving a conventional on-site inspection. In addition, the influence of decision-making (extrinsic) and system-related (intrinsic) cost assignments is examined to identify, at least qualitatively, cost levels that justify investment in SHM.

The rest of the paper is structured as follows: key theoretical principles used to quantify the VoI from thickness loss monitoring are presented in Section \ref{seq2}. The employed case study is described in Section \ref{seq3}. Section \ref{seq4} presents results from the numerical implementation of the framework and is followed by concluding remarks in Section \ref{seq5}.

\section{Quantifying the Value of Information for thickness loss monitoring}\label{seq2}

This section outlines the main elements used to quantify the Value of Information (VoI) for corrosion-induced thickness loss monitoring. First, a brief overview of Bayesian model updating is provided, focusing on the problem of interest. The VoI is then defined, and the decision setting for thickness loss monitoring used to formulate the consequence cost functions is introduced. Finally, the computational implementation of the proposed framework is summarised.

\subsection{Bayesian model updating for thickness loss monitoring}\label{seq2.1}

Let the quantity of interest (QoI) in this work—the corrosion-induced thickness loss at a specific region of the hull structure—be defined as a random variable (RV) $\Delta \mathcal{T} : \Omega_{\Delta \mathcal{T}} \rightarrow \mathbb{R}$. Here, $\Omega_{\Delta \mathcal{T}}$ denotes the sample space of the real-valued random variable, a scalar with realisations $\Delta \tau \in \mathbb{R}$. This definition implies uniform thickness loss over the region of interest. The focus is on monitoring thickness loss over time following its onset, i.e., the moment when the protective coating fails. This is reasonable, as tracking the evolution of deterioration enables timely maintenance decisions that can minimise downtime and control the costs associated with potential repairs.

A model is required to describe the temporal evolution of thickness loss. Let the operator describing this model be denoted as $\mathcal{F}$. Then, $\mathcal{F}\left( t ; \bm{\uptheta} \right)$ represents a stochastic model parametrised by a random vector $\bm{\uptheta} \in \mathbb{R}^{d_{\theta}}$, where $d_{\theta}$ denotes the number of model parameters. Deteriorating structural systems can be considered as a class of dynamical systems, and are therefore described using a stochastic state-space representation \citep{Kamariotis2023_v2}. In this setting, $\mathcal{F}\left( t ; \bm{\uptheta} \right)$ defines the state model, and $\Delta\tau$ is the state variable. To fully express the problem in this form, we write:

\begin{equation} \label{eq1}
\Delta \tau_k = \mathcal{F}\left( t_k ; \bm{\uptheta} \right) + \psi,
\end{equation}
\begin{equation} \label{eq2}
\mathbf{y}_k = \mathcal{H}\left( \Delta \tau_k ; \bm{\upphi} \right) + \bm{\upxi}.
\end{equation}

In Eqs. (\ref{eq1})--(\ref{eq2}), time is expressed in discrete form, with $\left\lbrace t_k \right\rbrace_{k=1}^{K}$ denoting the measurement instances. In Eq. (\ref{eq1}), $\psi$ represents the process noise—a random variable that quantifies state-model (epistemic) uncertainty. In Eq. (\ref{eq2}), the operator $\mathcal{H}$ denotes the observation (or measurement) model, which is in principle also stochastic and parametrised by a random vector $\bm{\upphi} \in \mathbb{R}^{d_{\phi}}$. This operator maps the state variable, which may not be directly observable, to the observation space. The system observations, represented by $\mathbf{y}_k \in \mathbb{R}^{N}$, are treated as noisy realisations of the observation model. The additive random vector $\bm{\upxi} \in \mathbb{R}^{N}$ denotes the observation noise and quantifies aleatoric uncertainty.

\begin{figure}[!t]
\centering
\includegraphics[scale=0.9]{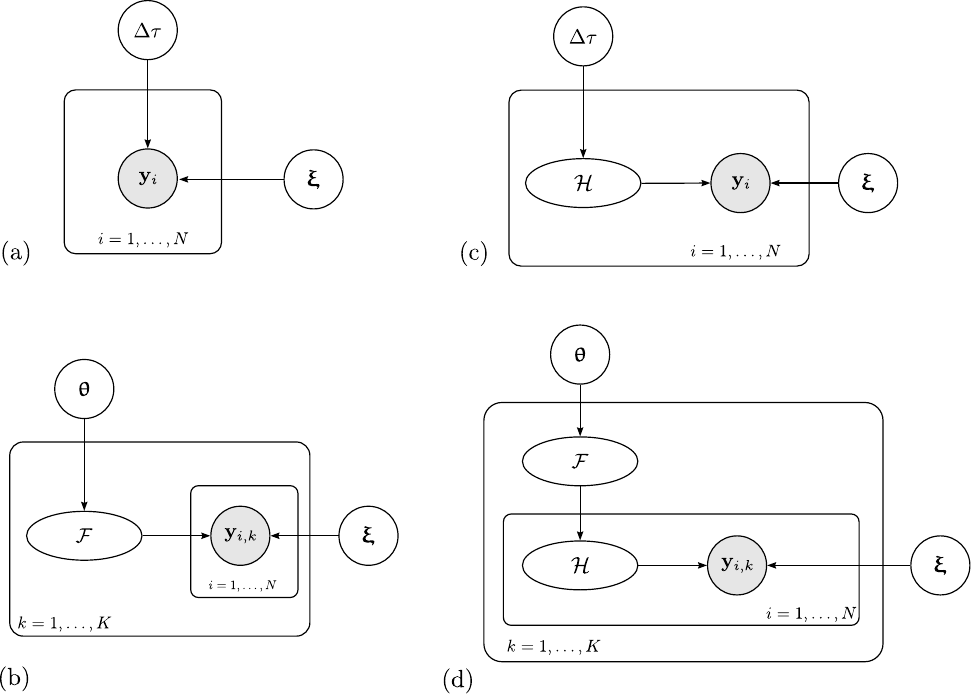}
\caption{Directed graphical models (DGMs) of Bayesian model updating problems considered for thickness loss monitoring.}
\label{Fig1}
\end{figure}

The state-space formulation described by Eqs. (\ref{eq1})--(\ref{eq2}) is well suited for monitoring applications, as it provides a general structure for defining Bayesian model updating (BMU) problems with different inferential QoIs and observation types. Figure~\ref{Fig1} presents the corresponding directed graphical models (DGMs) for the BMU problems considered in this work. Circular empty nodes denote inferred random variables (RVs); grey shaded nodes denote observed quantities; and ovals denote transformations of RVs. Arrows indicate conditional relationships, while plates iterating over $k$ (time) and $i$ (features) represent conditional independence.

Figures \ref{Fig1} (a)--(b) correspond to cases where the state variable is directly observable, i.e., the observation model $\mathcal{H}$ is the unitary operator. These models represent on-site inspections. In Figure~\ref{Fig1}~(a), the goal is to infer thickness loss at a specific time, so the state model $\mathcal{F}$ reduces to a probability density function (pdf) with known parameters $\bm{\uptheta}$. In Figure~\ref{Fig1}~(b), the objective is to infer the parameters of the deterioration model for lifecycle-level decision-making.

The DGMs in Figures~\ref{Fig1}~(c)--(d) depict cases where the state variable is not directly observable, e.g., in strain-based SHM. Here, the observation model is typically a structural response model parametrised by thickness loss, such as a finite element (FE) model or a data-driven surrogate. Figure~\ref{Fig1}~(c) represents inference of thickness loss at a specific time, where the state model becomes a prior distribution, whereas Figure~\ref{Fig1}~(d) concerns inference at the lifecycle level. In all DGMs in Figure~\ref{Fig1}, the observation noise term $\bm{\upxi}$ is also inferred from data. The probabilistic model used to describe this term, along with the corresponding quantity that is of interest for inference, is detailed in the discussion of Eqs. (\ref{eq3})--(\ref{eq4}).

The final DGM represents the general BMU problem, from which all other cases can be derived under the specific conditions described above. This general model is used hereafter as the reference formulation for all definitions. Two simplifications are introduced. First, since all analyses are numerical and the deterioration model is user-defined, the process noise term $\psi$ is neglected. Second, the observation operator $\mathcal{H}$ is assumed deterministic, and the associated parameter vector $\bm{\upphi}$ is therefore omitted. In this setting, the quantities of interest are collected in the random vector $\bm{\uptheta} \in \mathbb{R}^{d_{\theta}}$. Using Bayes’ theorem, the posterior distribution of $\bm{\uptheta}$ given the observations $\mathbf{y}$ is expressed as:
\begin{equation} \label{eq3}
p ( \bm{\uptheta} \vert \mathbf{y}) = \frac{p(\mathbf{y} \vert \bm{\uptheta}) p( \bm{\uptheta})}{\int_{\mathcal{D_{\bm{\uptheta}}}}p(\mathbf{y} \vert \bm{\uptheta}) p( \bm{\uptheta}) \, d\bm{\uptheta}},
\end{equation}
where \( p(\mathbf{y} \vert \bm{\uptheta}) \) denotes the likelihood function, \(p(\bm{\uptheta}) \) is the prior over the quantities of interest and the integral in the denominator is the Bayesian evidence term. The choice of prior may be either empirical or subjective, but ultimately, rests on the analyst.

For the problems considered here, the construction of the likelihood function relies on specifying a probabilistic model for the observation noise term $\bm{\upxi}$ in Eq.~(\ref{eq1}). This model, commonly referred to as the prediction error model \cite{Simoen2015}, characterises the statistical relationship between the observations and the model predictions. The prediction error is assumed to follow a zero-mean spherical Gaussian distribution, i.e., $\bm{\upxi} \sim \mathcal{N}\left(\mathbf{0}, \sigma^2 \mathbf{I}_{N}\right)$, where $\mathbf{I}_N$ is the $N$-dimensional identity matrix and $\sigma$ is the scalar standard deviation, also inferred from the data. The likelihood function is then written as:
\begin{equation} \label{eq4}
p ( \mathbf{y} \vert \bm{\uptheta} ) = \prod_{k=1}^{K}\prod_{i=1}^{N}\mathcal{N}(\mathbf{y}_{i,k} - \hat{\mathbf{y}}_{i,k};0 , \sigma^2),
\end{equation}
where $\hat{\mathbf{y}}_{i,k} = \mathcal{H}\left( \mathcal{F}\left(t_k; \bm{\uptheta} \right)\right)$, and stochastic independence has also been assumed for the observations at different points in time.

For the problems considered here, as in most practical applications, the likelihood and prior in the numerator of Eq.~(\ref{eq3}) do not form a conjugate pair. Consequently, the evidence term is intractable, and no analytical expression for the posterior is available. Sampling-based methods, namely Markov Chain Monte Carlo (MCMC), are therefore employed to approximate the true posterior distribution by drawing samples from it. Details of the specific algorithms used are provided in later sections on implementation.

\subsection{Defining the Value of Information}\label{seq2.2}

The Value of Information (VoI) arises from Bayesian decision analysis, in which the prior and posterior distributions play a central role. To define the VoI, we first consider the decision-making framework underpinning maintenance planning for deteriorating structural systems. The key quantity in this analysis is the state vector $\bm{\uptheta}$, which represents the decision-maker’s understanding of the level of deterioration experienced by the structure. Quantifying the VoI ultimately amounts to assessing whether an investment in acquiring additional information about the structure improves this understanding sufficiently to enable better-informed operation and maintenance decisions.

The process and mathematical formulation adopted here follow the definitions given by Chadha et al. \citep{Chadha2021}. The first step, known as the \textit{prior decision analysis}, closely resembles traditional engineering practice. In this setting, the decision-maker cannot obtain information about the current state of the structure, so maintenance decisions must rely solely on prior knowledge. This information may stem from historical data—such as past records from similar structures—or from expert judgement. Inevitably, the optimal maintenance decision is the one that minimises expected loss, which at a given time $t_k$ in the structure’s lifetime can be expressed as:
\begin{equation}\label{eq5}
d^{(k)}_{\text{prior}} = \underset{{d_j}}{\arg\min} \ \mathcal{C}^{(k)}_{\text{prior}} \left(d_j\right).
\end{equation}

In Eq.~(\ref{eq5}), $d_j \in \mathbf{d} = \left\lbrace d_0, \dots, d_D \right\rbrace$ denotes one of $D$ discrete maintenance decisions, and $\mathcal{C}^{(k)}_{\text{prior}}$ is the prior expected loss, or Bayes risk, at time $t_k$ in the structure’s lifetime, defined as:
\begin{equation} \label{eq6}
\mathcal{C}^{(k)}_{\text{prior}} \left( d_j \right) = \mathbb{E}_{p(\bm{\uptheta})} \left[\mathcal{R}\left( d_j, \bm{\uptheta} \right)\cdot\left( 1 + r_k  \right)^{t_k}  \right],
\end{equation}
where $\mathbb{E}_{p(\bm{\uptheta})}$ denotes the expected value operator with respect to the prior distribution and $\mathcal{R}\left(d_j, \bm{\uptheta} \right)$ refers to the consequence cost of making decision $d_j$.

The consequence cost function is expressed here in general terms. Its explicit definition and connection to the state variable are provided in the following sections. The term multiplying the consequence cost applies an inflation adjustment to define the cost at any future time $t_k$, when a decision is made. This adjustment, based on the annual inflation rate $r_k$, follows the formulation of Chadha et al.~\citep{Chadha2021}, which introduced the VoI definition adopted in this work.

The inflation rate is used instead of a discount rate, as it provides a simple and consistent means of representing cost trends. This choice is particularly suitable when concrete cost data are unavailable, as in this work. In contrast, defining a discount rate without such data requires subjective assumptions about factors such as opportunity cost or the time value of money, which—if made incautiously—can compromise result interpretability. Similar practices are observed in the literature: studies defining costs in a normalised (unitless) form typically employ inflation adjustments \citep{Chadha2021,Chadha2023,Chadha2025}, whereas those using concrete (monetary) costs adopt discount rates \citep{Zonta2014,Kamariotis2023}. The prior expected loss over the structure’s lifecycle can then be expressed as:
\begin{equation} \label{eq7}
\mathcal{C}_{\text{prior}} = \sum_{k=1}^{K}\mathcal{C}^{(k)}_{\text{prior}}\left( d_{\text{prior}}^{(k)} \right).
\end{equation}

The next step in quantifying the VoI is the \textit{pre-posterior decision analysis}, which is fundamental to VoI-based assessment during SHM system design. A \textit{posterior decision analysis}—the immediate extension of the prior—cannot be conducted at this stage, as SHM data are not yet available before system installation. Therefore, synthetic data must be generated to represent potential outcomes of the monitoring system.

In the context of SHM system design, it is reasonable to consider multiple candidate monitoring strategies, denoted by $\mathbf{z} = \left\lbrace z_1, \dots, z_{d_z} \right\rbrace$, where $d_z$ is the number of candidates. The corresponding observation vectors are indexed by strategy, i.e., $\mathbf{y}_{z_i}$, and are treated as random to capture uncertainty associated with data collection. As in the prior decision analysis, the optimal decision at any given time $t_k$ in the \textit{pre-posterior} analysis is that which satisfies:
\begin{equation}\label{eq8}
d^{(k)}_{\text{pre}}\left(z_i \right) = \underset{d_i}{\arg\min} \ \mathcal{C}^{(k)}_{\text{pre}} \left( d_j; \mathbf{y}_{z_i} \right).
\end{equation}

Here, $\mathcal{C}^{(k)}_{\text{pre}}\left(d_j; \mathbf{y}_{z_i}\right)$ denotes the conditional expected loss, or Bayes risk, associated with decision $d_j$ and the observation set $\mathbf{y}_{z_i}$ obtained from monitoring strategy $z_i$. Analogous to Eq.~(\ref{eq6}), the conditional expected loss expands to:
\begin{equation} \label{eq9}
\mathcal{C}^{(k)}_{\text{pre}} \left( d_j; \mathbf{y}_{z_i} \right) = \mathbb{E}_{p(\bm{\uptheta} \vert \mathbf{y}_{z_i})} \left[\mathcal{R}\left( d_j,\mathbf{\uptheta} \right)\cdot\left( 1 + r_k  \right)^{t_k}  \right].
\end{equation}

Eq.~(\ref{eq9}) represents an intermediate step in the \textit{pre-posterior} decision analysis, as it accounts for the updated knowledge obtained from a specific realisation of the monitoring data. Accordingly, the expectation is taken with respect to the posterior distribution of the state variable, $p\left(\bm{\uptheta} \mid \mathbf{y}_{z_i}\right)$, for a given outcome of monitoring strategy $z_i$.

To estimate the expected loss, or Bayes risk, over the structure’s lifecycle, uncertainty in the outcomes of the information acquisition strategy must also be considered. Using the optimal decision from Eq.~(\ref{eq8}), the expected \textit{pre-posterior} loss is obtained as the expected value of Eq.~(\ref{eq9}) with respect to the probability distribution of the observation vectors corresponding to the different monitoring strategies:
\begin{equation} \label{eq10}
\mathcal{C}^{(k)}_{\text{pre}}\left( z_i \right) = \mathbb{E}_{p\left( \mathbf{y}_{z_i} \right)} \left[ \mathcal{C}^{(k)}_{\text{pre}} \left( d^{(k)}_{\text{pre}}\left(z_i \right); \mathbf{y}_{z_i} \right) \right].
\end{equation}
 
Computing the expected \textit{pre-posterior} loss over the structure’s lifecycle for a given monitoring strategy requires accounting for the installation and operation and maintenance (O\&M) costs of the monitoring system. Assuming O\&M costs compound annually and the installation cost is paid at the present time, the expected lifecycle \textit{pre-posterior} loss for a given monitoring strategy is defined as:
\begin{equation}\label{eq11}
\mathcal{C}_{\text{pre}}\left( z_i \right) = \underbrace{\mathcal{C}(z_i) + \sum_{m=1}^{M}\mathcal{C}^{(m)}_{\text{O\&M}}(z_i)\cdot\left(1+r_m\right)^{t_m}}_{\text{intrinsic}} + \underbrace{\sum_{k=1}^{K}\mathcal{C}^{(k)}_{\text{pre}}(z_i)}_{\text{extrinsic}},
\end{equation}
where $M$ denotes the number of operational years for which the system is designed. The total cost comprises intrinsic (system-related) costs—installation and O\&M—and the extrinsic decision-making cost. The expected Value of Information (EVOI) is then intuitively defined as the difference between the expected losses obtained from the prior and \textit{pre-posterior} decision analyses, namely:
\begin{equation}\label{eq12}
\begin{split}
\mathrm{EVOI}\left( z_i \right) &= \mathcal{C}_{\text{prior}} - \mathcal{C}_{\text{pre}}\left( z_i \right) \\
& = \mathcal{C}_{\text{S}}\left( z_i \right) - \left( \mathcal{C}\left( z_i \right) + \mathcal{C}_{\text{O\&M}}\left( z_i \right) \right),
\end{split}
\end{equation}
where $\mathcal{C}_{\text{O\&M}}\left( z_i \right)$ is the total O\&M cost and $\mathcal{C}_{\text{S}}(z_i)$ \citep{Chadha2021} denotes the expected cost savings:
\begin{equation}\label{eq13}
\begin{split}
\mathcal{C}_{\text{S}}\left( z_i \right) = \sum_{k=1}^{K}\left\lbrace\mathbb{E}_{p(\bm{\uptheta})} \left[\mathcal{R}\left( d_{\text{prior}}^{(k)}, \mathbf{\uptheta} \right)\cdot\left( 1 + r_k  \right)^{t_k}  \right]\right\rbrace \\ 
- \sum_{k=1}^{K} \left\lbrace \mathbb{E}_{p\left( \mathbf{y}_{z_i} \right)} \left[ \mathbb{E}_{p(\bm{\uptheta} \vert \mathbf{y}_{z_i})} \left[\mathcal{R}\left( d^{(k)}_{\text{pre}}\left(z_i \right), \mathbf{\uptheta} \right)\cdot\left( 1 + r_k  \right)^{t_k} \right] \right] \right \rbrace.
\end{split}
\end{equation}

Essentially, Eq.~(\ref{eq13}) quantifies the value gained from acquiring monitoring information to support maintenance decisions. For an SHM system design to be feasible, this quantity must be strictly positive and exceed the system’s intrinsic costs to justify investment. Hence, feasible SHM designs are characterised by a positive EVOI. In this work, we adopt the alternative formulation proposed by Chadha et al.~\citep{Chadha2021}, known as the expected reward-to-investment risk ratio, defined as:
\begin{equation}\label{eq14}
\lambda(z_i) = \frac{\mathcal{C}_{\text{S}}(z_i)}{\mathcal{C}(z_i) + \mathcal{C}_{\text{O\&M}}\left(z_i\right)}.
\end{equation}

According to this definition, feasible designs are those for which $\lambda(z_i) > 1$. By effectively normalising the monetary units, this formulation introduces a more qualitative perspective and is well suited to this work, which lacks information on actual cost values. Finally, to compare different information acquisition strategies—not limited to SHM—we employ the relative risk-adjusted reward, defined as:
\begin{equation}\label{eq15}
\chi \left(z_1, z_2\right) = \frac{\lambda\left(z_2\right) - 1}{\lambda\left(z_1\right) - 1}.
\end{equation}

It is worth noting that, when this metric is applied to compare a candidate SHM strategy with a monitoring approach consisting solely of inspections, it yields—at least qualitatively—a measure analogous to the Value of SHM proposed by Kamariotis et al.~\citep{Kamariotis2023}.

\subsection{Defining a decision setting for thickness loss monitoring}\label{seq2.3}

The expected costs in Eqs.~(\ref{eq6}), (\ref{eq9}), and (\ref{eq10}) follow the definition of Chadha et al.~\citep{Chadha2021}, where the relevant consequence cost functions were expressed directly in terms of the state variable. In this work, we deviate from that approach and employ a derivative quantity, which still depends on the probability distribution, or realisations, of the state variable. Specifically, we use an exceedance probability estimate defined for a particular threshold value of the state variable, i.e., thickness loss.

This choice provides a more informative basis—from the perspective of the decision-maker—on which to define consequence costs for maintenance decisions. In practice, decision-makers tend to focus on threshold values or acceptable limits rather than direct estimates of the quantity of interest. Moreover, this formulation implicitly captures risk perception through the selection of the threshold value.

Let a sample of thickness loss realisations $\left\lbrace \Delta\tau_n (t_k) \right\rbrace_{n=1}^{N_{\text{CITL}}}$ be available at time $t_k$. These realisations may originate from either a prior or posterior distribution and may have been obtained directly or through forward uncertainty propagation of the state model parameters, according to Eq.~(\ref{eq1}). We further define a random variable to describe the threshold value, $\Delta \mathcal{T}^{\text{(th)}} : \Omega_{\Delta \mathcal{T}^{\text{(th)}}} \rightarrow \mathbb{R}$, which is described by its probability density function (pdf) $p\left( \Delta \mathcal{\tau}^{\text{(th)}} \right)$. This quantity is treated as random to represent the epistemic uncertainty associated with the assumption of uniform thickness loss, which rarely holds in practice. This probabilistic treatment implicitly accounts for spatial variability by capturing uncertainty in the representative threshold value.

The pdf $p\left( \Delta \mathcal{\tau}^{\text{(th)}} \right)$ must be selected to appropriately represent the epistemic uncertainty modelling assumption. Its shape should avoid introducing unnecessary bias into the analysis---for instance, heavy-tailed distributions should be avoided. Its parameters must also ensure that the support corresponds to physically meaningful values of the thickness loss threshold. This can be achieved either explicitly by the choice of probability model or implicitly through sampling considerations. The \textit{interval} exceedance probability at time $t_k$ is then defined as:
\begin{equation}\label{eq16}
\hat{\text{P}}^{\text{(i)}}_{\text{ex}} (t_k) = \mathbb{P} \left[\Delta \mathcal{T} (t_k) > \Delta \mathcal{T}^{(\text{th})} \right] \approx \frac{1}{N_{\text{CITL}}}\sum_{n=1}^{N_{\text{CITL}}} \ind \left[ \Delta \tau_n (t_k) > \Delta \tau_{n}^{(\text{th})}  \right],
\end{equation}
where $\ind[\cdot]$ denotes the indicator function and $\Delta \tau_{n}^{(\text{th})}$ is a realisation drawn from $\Delta \mathcal{T}^{\text{(th)}}$. The estimator in Eq. (\ref{eq16}) is a standard Monte Carlo (MC) estimator; therefore, the number of posterior samples $N_{\text{CITL}}$ must be sufficiently large for the estimated probabilities to converge to their true values. Further discussion about the convergence characteristics of this estimator is provided in later sections.

The quantity $\hat{\text{P}}^{\text{(i)}}_{\text{ex}} (t_k)$ expresses the probability that thickness loss at a given time has exceeded a specified threshold value. By definition, it does not account for potential exceedance events occuring before time $t_k$. Accordingly, a cumulative exceedance probability is used as the attribute for defining the consequence functions. This probability, denoted as $p^{(k)}_{\text{ex}}$, is defined as the complement of the probability of non-exceedance up to time $t_k$, assuming independent non-exceedance events, and is expressed as:
\begin{equation}\label{eq17}
p_{\text{ex}}^{(k)} = \hat{\text{P}}^{\text{(c)}}_{\text{ex}}(t_k) = 1 - \prod_{j=1}^{k}\left[ 1 - \hat{\text{P}}^{\text{(i)}}_{\text{ex}}(t_j) \right].
\end{equation}

\subsection{Defining consequence cost functions} \label{seq2.4}

To define the consequence cost functions, we must first specify the decision setting, i.e., determine the characteristics of the decision set $\mathbf{d}$. Here, a simple binary decision setting is adopted, hence the decision set $\mathbf{d} = \left\lbrace d_0, d_1 \right\rbrace$, where $d_0$ denotes the decision not to repair and $d_1$ denotes the decision to repair. In this framework, maintenance decisions are triggered by threshold exceedance events, rather than structural failure in the classical sense (yield, buckling, ultimate strength). The consequence costs quantify the economic impact of these maintenance decisions.

\begin{figure}[!t]
\centering
\includegraphics[scale=0.8]{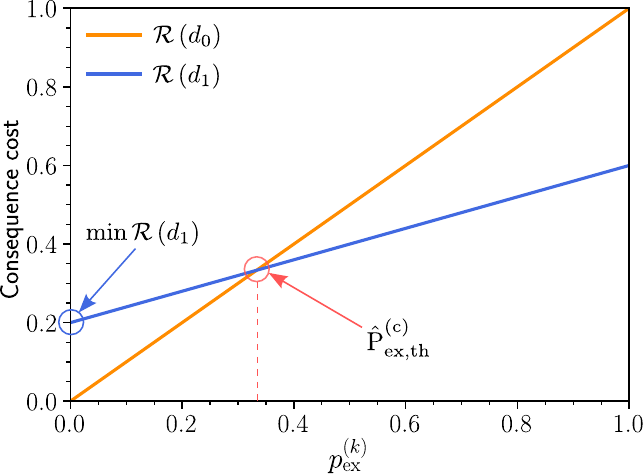}
\caption{Consequence cost functions for $\mathbf{d} = \left\lbrace d_0, d_1 \right\rbrace$.}
\label{Fig2}
\end{figure}

Accordingly, consequence cost functions are defined for each decision and parametrised with respect to the cumulative exceedance probability. A linear formulation is employed in this work, corresponding in principle to a risk-neutral approach. The cost functions are illustrated schematically in Figure \ref{Fig2}.

Consequence costs are defined in normalised form, between 0 and 1, due to the limited cost information available for the case study considered. This normalisation enables the cost extrema to be defined rationally and facilitates a straightforward interpretation of the relative relationships between different costs.
Accordingly, the maximum overall cost is set to 1 and corresponds to the case where the decision not to repair is made while the cumulative exceedance probability with respect to the thickness loss threshold approaches unity. Conversely, the minimum cost value is set to 0 and applies when the decision not to repair is made and $p^{(k)}_{\text{ex}} \rightarrow 0$. The corresponding consequence cost function is therefore defined as:
\begin{equation} \label{eq18}
\mathcal{R} \left( d_0, p_{\text{ex}}^{(k)} \right) = p_{\text{ex}}^{(k)}.
\end{equation}

The consequence cost function for the decision to repair, $d_1$, is defined using two parameters that serve as indicators of the decision-maker's risk profile. The first, denoted $\min \mathcal{R}(d_1)$, represents the cost of assigning a repair when the probability of exceeding the maintenance-related threshold is near-zero. This corresponds to the false alarm cost and reflects the decision-maker’s tendency towards risk aversion or risk seeking.

The second parameter is a threshold value of the exceedance probability, denoted as $\hat{\text{P}}^{\text{(c)}}_{\text{ex,th}}$, above which the decision-maker always opts to repair. This parameter also implicitly models the decision-maker's risk perception.  By definition, it corresponds to the intersection point of the two cost functions, leading to the following expression for the decision-to-repair consequence cost:
\begin{equation}\label{eq19}
\mathcal{R} \left( d_1, p_{\text{ex}}^{(k)} \right) = \min \mathcal{R}(d_1) + \left[1 - \frac{\min \mathcal{R}(d_1)}{\hat{\text{P}}^{\text{(c)}}_{\text{ex,th}}} \right]\cdot p_{\text{ex}}^{(k)}.
\end{equation}

It is important to acknowledge that, while the consequence cost functions defined here incorporate an element of risk perception modelling, they do not constitute formal utility functions in the decision-theoretic sense. They may therefore be viewed as occupying a middle ground between a risk-neutral approach and the more comprehensive treatments presented by Chadha et al. \citep{Chadha2023} or Mir Rangrez et al. \cite{MirRangrez2025}.

Furthermore, it is useful to clarify the distinction between the present formulation and previous Value of Information (VoI) frameworks employing reliability-based metrics. Although exceedance probabilities and related measures have been used in structural reliability-based decision analyses, our approach differs in that consequence costs are defined directly as functions of the exceedance probability $p_{\text{ex}}^{(k)}$ (Eqs.~\ref{eq18}--\ref{eq19}), rather than being conditioned on the state variable and integrated with respect to its distribution to compute expectations.

This formulation enables decision-makers to specify maintenance-relevant thresholds, $\Delta\tau^{(\text{th})}$, explicitly, thereby embedding risk perception and operational constraints directly into the cost structure—a feature particularly suited to practical maintenance planning contexts.

\subsection{Computational implementation}\label{seq2.5}

\begin{figure}[!htp]
\centering
\includegraphics[scale=1.0]{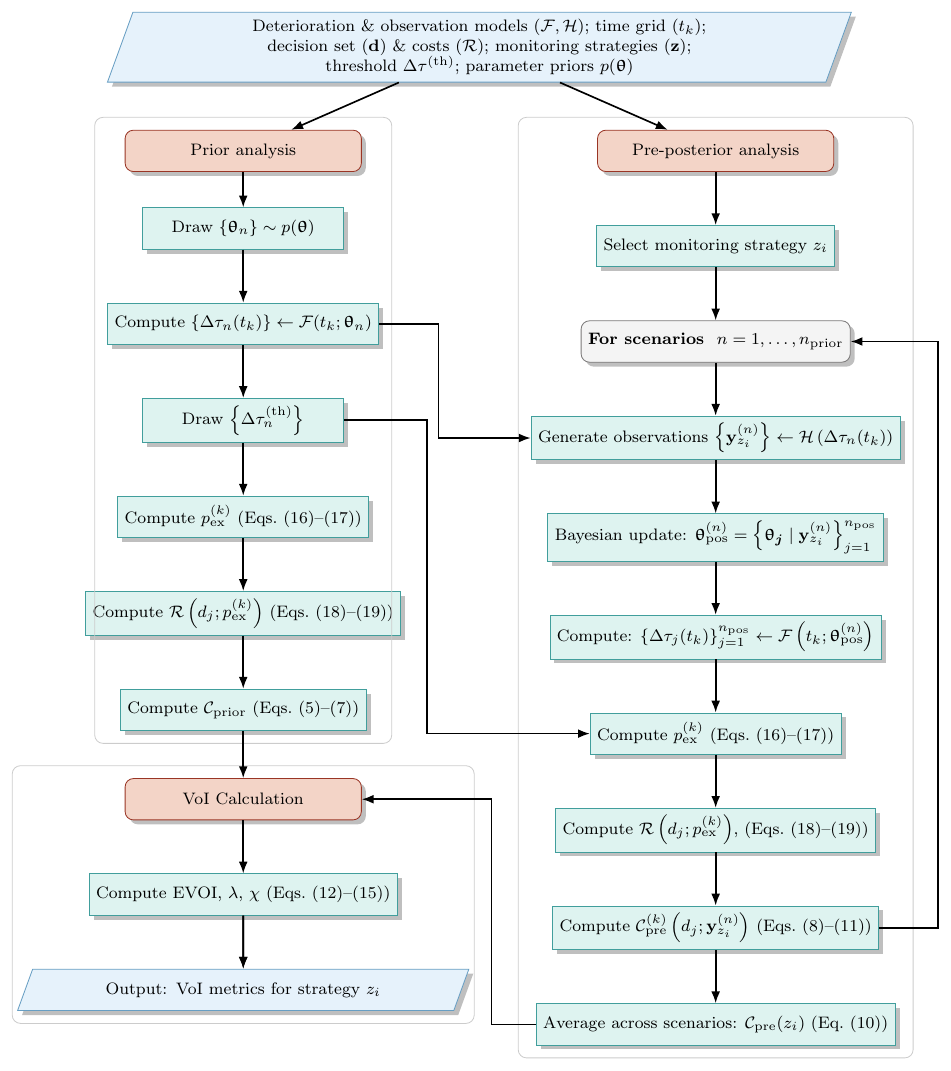}
\caption{Flowchart illustrating the computational workflow of the proposed VoI-based assessment framework for strain-based thickness loss monitoring in ship hull structural maintenance planning.}
\label{algo_flow}
\end{figure}

In the previous sections, the expected loss (Bayes risk) was introduced formally; this section outlines its computational implementation within the VoI framework. Figure~\ref{algo_flow} summarises the procedure through a flowchart distinguishing the prior and pre-posterior analyses.

The left branch represents the prior decision analysis, including sampling parameter priors, forward propagation to generate thickness loss realisations, and prior cost evaluation. The right branch describes the pre-posterior analysis, covering observation generation for a given monitoring strategy, Bayesian updating, and posterior cost computation. The two branches converge at the VoI calculation step, where the relevant metrics—EVOI, $\lambda$, and $\chi$—are obtained.

Sampling methods are employed for Bayesian updating in all Bayesian model updating problems considered here, as well as for evaluating the expected values required in the VoI calculation. Crude Monte Carlo (MC) estimators, analogous to Eq.~(\ref{eq16}), are used for this purpose but are not written explicitly to avoid unnecessarily dense notation. Implementation details, including sample sizes, are provided in subsequent sections.

\clearpage

\section{Case study description}\label{seq3}

This section describes the case study used to demonstrate the proposed framework. It includes the definition of the stochastic deterioration model and the description of the different monitoring strategies and their associated observation models.

\subsection{Probabilistic thickness loss modelling}\label{seq3.1}

Pre-posterior decision analysis requires generating realisations of the deterioration process. These realisations represent potential deterioration histories that the structure may experience during its lifetime. Here, a deterioration model with known parametric structure is assumed, characterised by the parameter vector $\bm{\uptheta}$. Specifically, the temporal evolution of thickness loss is described as:
\begin{equation}\label{eq20}
\Delta\tau_k = \mathcal{F} \left(t_k ; \bm{\uptheta} \right) = \frac{\gamma}{\alpha + 
 \beta\exp{\left( -\left( t_k - t_0 \right) \right)}},
\end{equation}
where $\bm{\uptheta} = \left\lbrace \alpha, \beta, \gamma \right\rbrace$ and $t_0$ denotes the time of corrosion initiation, assumed here to be the 10$^{\text{th}}$ year of vessel operation.
The logistic-type model in Eq. (\ref{eq20}) follows Qin et al. \citep{Qin2003}. However, in this work the parameter $\gamma$, which controls the asymptotic maximum thickness loss, is inferred from data, whereas in the original formulation it represents the known maximum observed thickness loss. The remaining  parameters, $\alpha, \beta$, jointly govern the initial corrosion rate and the temporal evolution of the deterioration process.

This model yields thickness loss realisations consistent with historical data and the commonly understood physics of corrosion evolution. Specifically, it exhibits asymptotic behaviour typical plate-like components subjected to uniform corrosion, as the accumulation of corrosion by-products on the surface leads to a gradual deceleration of the process.


To generate realisations from the thickness loss prior process, we must first define the joint prior distribution over the deterioration model parameters $p \left( \bm{\uptheta} \right)$. The parameters are assumed to be independent random variables, each following a probability distribution chosen to produce physically consistent thickness loss realisations. The selected prior distributions are given below:
\begin{equation} \label{eq21}
\begin{split}
\alpha & \sim \mathcal{U}\left( 4.,  13. \right),  \\
\beta & \sim \mathcal{N}\left( 250., 50. \right), \\ 
\gamma & \sim \mathcal{U}\left( 4., 8.5 \right).
\end{split}
\end{equation}

\begin{figure}[!t]
\centering
\includegraphics[scale=1.0]{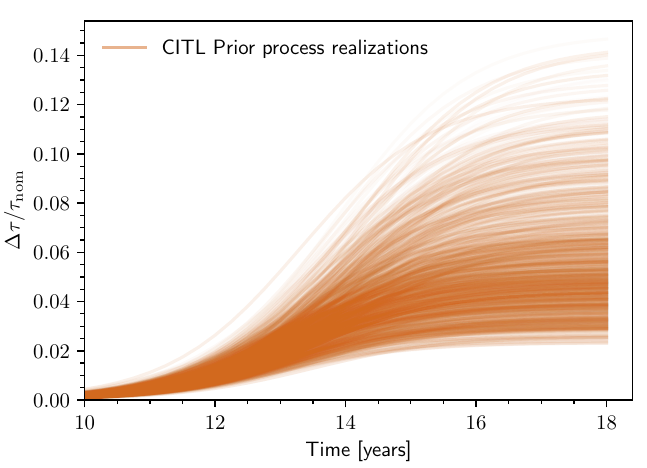}
\caption{Thickness loss prior process realisations expressed as a fraction of the as-built plate thickness $\tau_{\text{nom}} = 14 \ \text{mm}$.}
\label{Fig3}
\end{figure}

The choice of these priors can be illustrated through the thickness loss realisations shown in Figure~\ref{Fig3}. These were obtained by drawing $n_{\text{prior}} = 1000$ independent samples from each distribution in Eq.~(\ref{eq21}) and propagating them through the deterioration model. An eight-year period, beginning at $t_0 = 10$~years, was assumed during which corrosion is active and simultaneously monitored.

The parameter priors were selected through iterative refinement to ensure, together with the model structure, that the resulting thickness loss realisations satisfy several physical constraints. First, this choice yields strictly positive, monotonically increasing thickness loss time-histories. It also ensures that maximum thickness loss levels remain within realistic bounds, consistent with practical experience—specifically, below 20\% of the as-built plate thickness, a typical threshold for regulatory plate renewal. Finally, these priors produce a relatively wide prior process, desirable from a mathematical standpoint as it prevents the introduction of unnecessary bias into the model.

\subsection{Defining monitoring strategies}\label{seq3.2}

The generated thickness loss realisations are used to obtain corresponding realisations of the observation vectors for the different monitoring strategies, for which the Value of Information (VoI) will be quantified. These observation vector realisations are then used to evaluate the uncertainty associated with information acquisition through the Monte Carlo (MC) estimator of the expectation in Eq.~(\ref{eq10}) (see also Figure~\ref{algo_flow}). The first step in this process is to define the monitoring strategies considered in this work. The following cases are examined:
\begin{itemize}
\item \textit{Inspection only} $(z_0)$: A single inspection event takes place at $t_{\text{insp}} = 15$~years, during which thickness loss is directly observable.
\item \textit{Strain-based thickness loss identification} $(z_1)$: Strain data are used to identify the level of thickness loss directly at different points in time.
\item \textit{Strain-based thickness loss monitoring (limited data)} $(z_2)$: Strain data are used to infer the parameters of a thickness loss deterioration model, with a limited number of observations available.
\item \textit{Strain-based thickness loss monitoring} $(z_3)$: The same scenario as $z_2$, but with a larger number of available observations.
\end{itemize}

It is important to clarify the rationale behind each candidate monitoring strategy and its relevance to maintenance planning. The inspection-only strategy provides a baseline for comparison with current corrosion inspection practices, with the inspection time chosen to align with a five-year hull survey. The remaining three strategies correspond to typical SHM scenarios, where monitoring occurs more frequently—though not necessarily continuously—and deterioration is monitored indirectly through its effect on the structural response.
								
These SHM scenarios are differentiated by two factors: the monitored quantity and the amount of available data. The first reflects different decision-making biases; for instance, the decision-maker in $(z_1)$ does not assume a specific functional form for the deterioration model, unlike in the other scenarios. This distinction represents a differing attitude toward introducing inductive bias into the model. The second factor reflects varying levels of investment in intrinsic system costs, since acquiring and storing larger volumes of data indicate a greater willingness to commit resources.

\subsection{Defining observation models}\label{seq3.3}

An additional technical factor that differentiates the candidate monitoring strategies is the observation model required to generate observation vector realisations and construct the Bayesian likelihood function of Eq. (\ref{eq4}). In the inspection-only case, where thickness loss is directly observable, the observation model $\mathcal{H}$ is represented by a probabilistic model---a probability distribution. In contrast, the strain-based SHM scenarios require a structural model that maps thickness loss to strain at specific locations of interest.

The physical basis for this mapping is that thickness loss reduces the section modulus and flexural rigidity of structural components, thereby decreasing their stiffness. Under applied loading, this stiffness reduction manifests as increased strain at monitored locations, providing an indirect but observable indicator of thickness loss.

\begin{figure}[!t]
\centering
\includegraphics[scale=0.90]{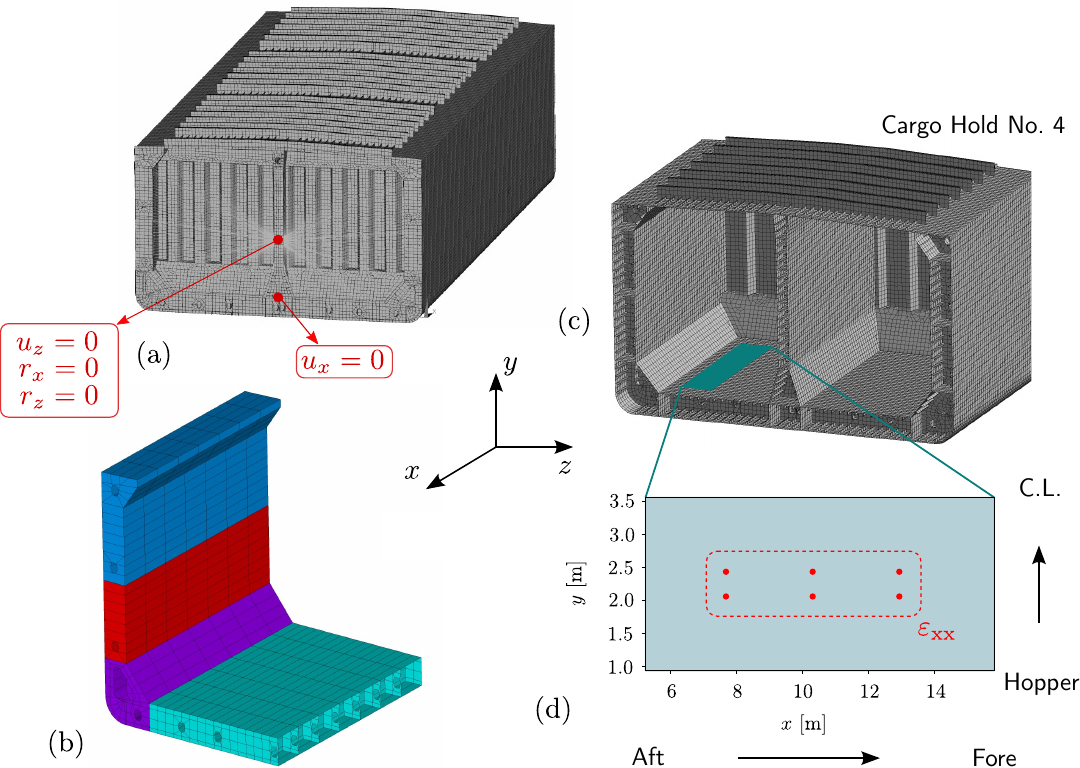}
\caption{(a) Perspective view of three hold compartment FE model, (b) regions with different thickness levels, (c) interior view of Cargo Hold 4  \& (d) longitudinal strain sensor locations expressed in local coordinates.}
\label{Fig4}
\end{figure}

The structural system employed as the case study to demonstrate the proposed framework is part of a ship hull structure. Specifically, it consists of the three central holds of a product carrier vessel with length $L_{\text{B.P.}} = 174.0 \ \text{m}$, beam $B = 32.2 \ \text{m}$ and depth $D = 19.1 \ \text{m}$. A perspective view of the meshed FE model of this structure is shown in Figure \ref{Fig4} (a), while Figure \ref{Fig4} (d) provides a zoomed-in view of the monitoring region. This region is located on the inner bottom of Cargo Hold No. 4 (C.H. 4)---the central of the three holds—and is shown in greater detail in Figure \ref{Fig4} (c). It represents the area where uniform thickness loss is inferred under the four monitoring strategies defined previously.

The level of thickness loss is considered constant, percentage-wise, across all structural components in the highlighted region of Figure \ref{Fig4} (b). Figure \ref{Fig4} (b) depicts one symmetric half of the hold, which includes the region of monitoring interest corresponding to the water ballast tank in the double bottom of the vessel. Since environmental conditions (salinity, temperature) are relatively similar within each ballast tank, a constant corrosion rate—and thus a uniform level of thickness loss—is assumed for the corresponding region.

This uniform assumption is appropriate for the monitoring objective here: monitoring thickness loss to inform maintenance scheduling. In this context, the global strain response is governed by overall stiffness reduction rather than localised variations such as pitting, which primarily affect local stress concentrations and contribute to different deterioration modes.

Regions with different colours in Figure~\ref{Fig4}~(b) correspond to the other ballast tanks within the cargo hold. Each is assumed to experience a constant level of thickness loss, distinct from that in the double bottom. Although not directly inferred, these thickness loss levels still influence the structural response as latent variables.

From a monitoring perspective, the inspection-only strategy $(z_0)$ is unaffected by these latent parameters, since observations are gathered directly from the region of interest. The probabilistic model used to generate these observations can be interpreted as accounting for both measurement uncertainty, as well as spatial variability. We note, however, that more expressive approaches exist for modelling the spatial variability of corrosion-induced thickness loss, such as representing it as a random field \citep{Georgiadis2021}.

However, incorporating such approaches would entail significant increases in computational cost and greater model complexity, due to the introduction of additional random field parameters and the need for refined FE meshes to capture local variations. Nevertheless, the global strain response that underpins the monitoring strategies evaluated here is primarily sensitive to overall stiffness reduction rather than local thickness variations. The Value of Information framework itself could accommodate non-uniform corrosion scenarios through reformulation of the quantity of interest and exceedance probability definitions. Such a case would instead constitute a different monitoring objective focused on local deterioration modes.

As a result, such approaches were considered beyond of the scope of this paper, and the following probabilistic model was used to generate inspection data:
\begin{equation}\label{eq22}
\mathrm{y}_{\text{insp}} \sim \mathcal{N}\left( \Delta \tau \left( t_{\text{insp}}; \bm{\uptheta}_n \right), \left( 0.1 \cdot \Delta \tau \left( t_{\text{insp}} ; \bm{\uptheta}_n \right) \right)^2 \right),
\end{equation}
which implies that, for every prior realisation $\bm{\uptheta}_n$, the inspection observations are normally distributed with a mean equal to the prior process value at the inspection time and a coefficient of variation (CoV) of 0.1. The probabilistic model in Eq. (\ref{eq22}) refers to a scalar random variable, from which $N$ samples are drawn to form the observation vector $\mathbf{y}_{\text{insp}} = \left\lbrace \mathrm{y}^{(i)}_{\text{insp}} \right \rbrace_{i=1}^{N}$.

For the strain-based monitoring strategies, the observation model is based on the finite element (FE) model of the three compartments presented in Figure \ref{Fig4}. The model is constructed using 4-node linear shell elements for all plate-like structural components and 2-node linear beam elements for all longitudinal stiffeners and flanges attached to primary supporting members, e.g., web frames. An element size of approximately 0.5 m was used for a model spanning 63.2 m lengthwise, resulting in a generally uniform mesh except in regions requiring refinement, such as around openings.

Boundary conditions were applied at the model ends using kinematic constraints implemented through master-slave connections. Specifically, the degrees of freedom shown in Figure \ref{Fig4} (a) were constrained at the master nodes, and these constraints were transferred to the slaves via rigid link elements. The slave nodes consist of all nodes attached to longitudinally extending structural members. Longitudinal displacement was constrained at a single node on the longitudinal plane of symmetry of the model to ensure equilibrium.

The model was subjected to external (hydrostatic) and internal (cargo) pressure loads, the latter treated as stochastic. These loads represent a typical loading condition for the vessel type considered in this work---namely, the homogeneous loading condition. A static, linear elastic analysis was performed, which is representative of still-water conditions---sailing in calm seas or being in port. This choice reflects the slow timescales of deterioration evolution, and therefore maintenance decision-making, which justify neglecting dynamic and wave-induced loads in favour of static loading scenarios that are sufficient for the monitoring objective (see also Section \ref{seq4.1}).

The stochastic treatment of cargo filling rates captures operational variability and enables generation of realistic strain observations that reflect the influence of the human factor in the cargo loading process. Specifically, cargo hold filling rates for the homogeneous loading condition were treated as random variables following a beta distribution, which are independent and identically distributed across cargo holds. More details on the applied loads and FE modelling can be found in Silionis et al. \citep{SilionisMS} and are omitted here for brevity.


The FE model used to generate strain observations has not been validated against strain measurements from operating ship structures, as such data are unavailable publicly. However, the model has been partially validated against analytical and semi-analytical methods for the different stress response levels commonly considered for ship structures, as documented in \citep{SilionisMS}. These include primary stresses due to global hull girder bending, secondary stresses caused by plate-stiffener response to lateral pressure, and tertiary stresses resulting from local plate bending. Moreover, the model follows industry-standard procedures consistent with IACS Common Structural Rules.

The absence of data from proposed sensor locations is inherent to pre-posterior analysis, since the methodology evaluates whether deployment is justified before installation. The approach presented here demonstrates a methodological framework built upon a high-fidelity, physics-based testbed, which can readily incorporate improved models or operational data as they become available. While quantitative Value of Information results are case-specific, relative comparisons between monitoring strategies are expected to remain valid under moderate model-related uncertainty.

The placement of potential strain sensors is of primary importance in this work, as it ultimately defines the observation model. Six sensor locations measuring longitudinal strain, $\varepsilon_{\text{xx}}$, were selected, arranged in an orthogonal grid, as shown in Figure \ref{Fig4} (d). Their placement and the measured strain component were determined using expert knowledge. Specifically, the sensors are located at the centres of the plate elements defined between the longitudinal stiffeners and transverse webs that make up a typical stiffened panel. These locations exhibit relatively high strain values, providing robustness against noise, and are more sensitive to the stiffness reduction caused by thickness loss, due to being located away from stiffer supporting members. For further details on this selection, the reader is referred to the earlier work of the authors \citep{SilionisMS}.

\begin{figure}[!b]
	\centering
	\includegraphics[scale=0.75]{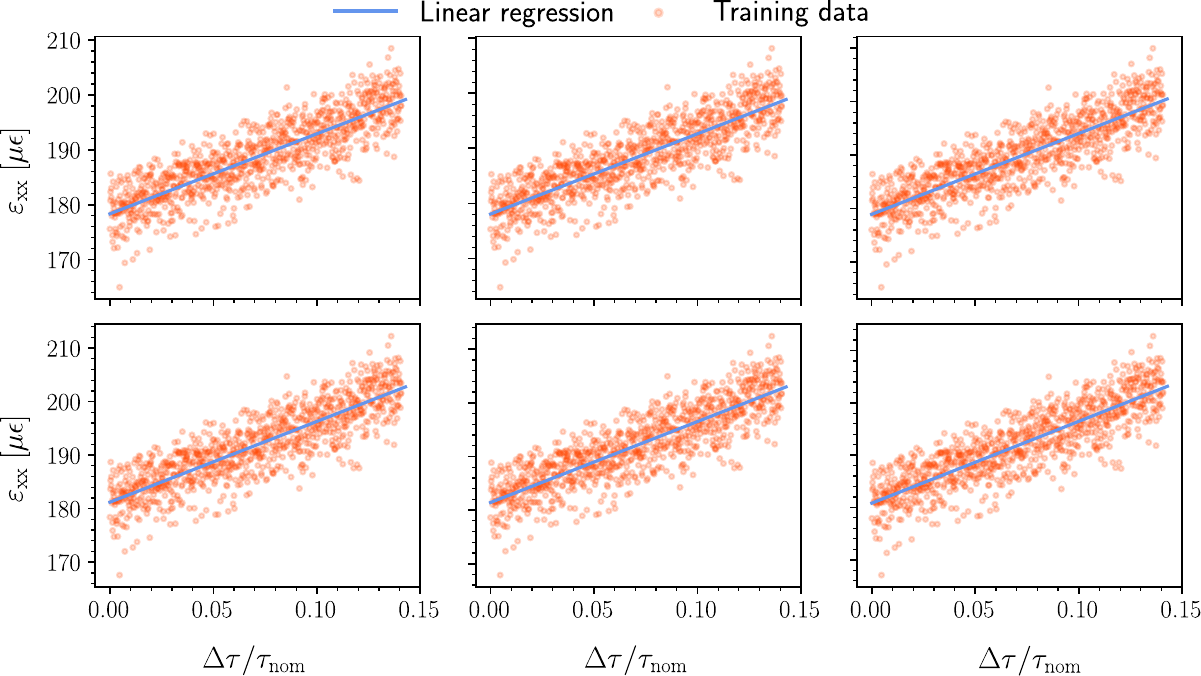}
	\caption{FE-based strain observations at potential sensor locations alongside linear regression-based observation models (figures follow the layout of Figure \ref{Fig4} (d)).}
	\label{Fig5}
\end{figure}

The scatter points in Figure \ref{Fig5} represent longitudinal strain data obtained at these locations for different thickness loss levels. The strain data were generated using a non-intrusive forward Monte Carlo process to propagate uncertainty related to both loading and deterioration. The latter refers to the variability in thickness loss affecting regions neighbouring the double bottom. Thickness loss is implemented in the FE model through proportional reduction of plate thickness in the affected regions, achieved by modifying the properties of the corresponding shell and beam elements, thus preserving the mesh topology.

Different scenarios are evaluated by parametrically varying the thickness reduction percentage and re-solving the static analysis in a non-intrusive manner, thereby generating the observation model training data shown in Figure \ref{Fig5}. The dispersion exhibited by the strain data is a result of the two sources of uncertainty mentioned earlier. In total, 1000 data points were generated by solving the FE model non-intrusively for different thickness loss and load realisations, requiring approx. 16 hours of wall time on a workstation equipped with 32 GB of RAM and an Intel\textregistered \ Core$^{\text{TM}}$ i7-10750H CPU.

The time required for a single FE model solution scales poorly with the computational demands of Markov Chain Monte Carlo (MCMC) algorithms. Each likelihood evaluation requires a call to the FE solver to compute the observation model, and such evaluations typically number in the tens of thousands. When extended to a pre-posterior decision analysis, in which MCMC is performed for each prior realisation, the computational cost increases by an additional order of magnitude. Consequently, a surrogate of the FE model must be trained to serve as the observation model $\mathcal{H}$ for the strain-based monitoring strategies.

A deterministic observation model is adopted in this work, although latent variables—stemming from load- and deterioration-related uncertainty---affect the strain responses. Using the mean function of probabilistic surrogate models is common practice (see e.g. \citep{Cristiani2021,Ramancha2022}). In contrast, employing a fully probabilistic observation model requires a marginalised likelihood function formulation to integrate out nuisance (latent) variables. For the problem in question, the authors found that this approach yields only marginal improvements in posterior quality while increasing wall time by a factor of four. Further details on this investigation are provided in Silionis et al. \citep{SilionisEWSHM} and are omitted here for brevity.

A simple linear regression model was ultimately sufficient to capture the mean trend of the strain data, as shown in Figure \ref{Fig5}. Bayesian inference (MCMC) was used to obtain posterior distributions of the regression parameters. Although the full posteriors were not used for this particular analysis, employing a probabilistic surrogate model maintains consistency with the general principles outlined earlier, and with the framework proposed by the authors in \citep{SilionisEWSHM}.

It is worth commenting on certain characteristics of the strain data. Similar strain patterns are observed across all sensor locations, indicating redundancy in the observations—an expected outcome given the comparable structural arrangement and loading conditions. This redundancy can, however, be beneficial in mitigating the influence of measurement noise. For higher-dimensional data, feature extraction techniques such as Singular Value Decomposition (SVD) could be applied to reduce dimensionality while preserving the essential statistical characteristics of the features.

\section{Numerical investigation}\label{seq4}

This section presents results from the implementation of the proposed Value of Information (VoI) quantification framework. Table \ref{tab:prob_models} summarises the probabilistic models and parameters used in the analysis, together with their locations in the text. Inference results are shown for the Bayesian model updating (BMU) problems corresponding to each monitoring strategy, followed by an examination of the behaviour of the employed limit state function. The influence of the decision threshold and different cost assignments on the VoI is then investigated. Finally, the relative utility of collecting inspection data is discussed.

\subsection{Bayesian inference of corrosion-induced thickness loss}\label{seq4.1}

The BMU problems considered in the proposed framework can only be solved using sampling methods, the most popular of which is arguably Markov Chain Monte Carlo (MCMC). MCMC refers to a family of algorithms that generate correlated samples from the target Bayesian posterior by constructing ergodic Markov chains with that posterior as their invariant distribution \citep{Brooks2011}. In this work, Hamiltonian Monte Carlo (HMC) was used in the form of the No U-Turn Sampler (NUTS) \citep{Hoffman2011}, a variant of HMC  designed for efficient exploration of high-dimensional parameter spaces. NUTS offers notable computational efficiency and rapid convergence to the invariant distribution, making it well suited to the demands of pre-posterior decision analysis, where numerous MCMC runs are required and computational resources must be used efficiently.

There is a rich literature on HMC in general, and NUTS in particular, such as the comprehensive review of Betancourt \citep{Betancourt2017}. A detailed description is beyond the scope of this paper, and the interested reader is referred to that work and the references cited earlier \citep{Brooks2011,Hoffman2011}. In this study, NUTS was implemented using the Python probabilistic programming library NumPyro \citep{Bingham2019,Phan2019}, which employs automatic differentiation to compute the gradients of the Bayesian likelihood function required by NUTS—gradients that are analytically intractable for the present problem.

\begin{table}[!b]
\centering
\caption{Summary of probabilistic models and parameters used in this work.}
\label{tab:prob_models}
\begin{tabular}{clc}
\toprule
\multicolumn{1}{c}{\textbf{Parameter/Model}} &
\multicolumn{1}{c}{\textbf{Distribution/Value}} & 
\multicolumn{1}{c}{\textbf{Location}} \\
\midrule
\multicolumn{3}{c}{\textit{Deterioration Model Parameters (Prior Distributions)}} \\
\midrule
$\alpha$ & Uniform: $\mathcal{U}(4, 13)$ & Eq. (\ref{eq21}) \\
$\beta$ & Normal: $\mathcal{N}(250, 50)$ & Eq. (\ref{eq21}) \\
$\gamma$ & Uniform: $\mathcal{U}(4, 8.5)$ & Eq. (\ref{eq21}) \\
$\Delta \tau_{k}$ & Uniform: $\mathcal{U}(0, 2.0) \ \mathrm{mm}$ & Sec. \ref{seq4.1} \\
\midrule
\multicolumn{3}{c}{\textit{Decision Framework Parameters}} \\
\midrule
$\Delta\mathcal{T}^{(\text{th})}$ & Normal: $\mathcal{N}(\mu, \sigma^2)$ (User-defined parameters) & Sec. \ref{seq2.3} \& \ref{seq4.2} \\
\midrule
\multicolumn{3}{c}{\textit{Observation Models}} \\
\midrule
Inspection data: $\mathrm{y}_{\text{insp}}$ &  Normal: $\mathcal{N}\left( \Delta \tau \left( t_{\text{insp}}; \bm{\uptheta}_n \right), \left( 0.1 \cdot \Delta \tau \left( t_{\text{insp}} ; \bm{\uptheta}_n \right) \right)^2 \right)$ & Eq. (\ref{eq22}) \\
Prediction error st. dev.: $\sigma$ & Half-Normal(1.0) & Sec. \ref{seq4.1} \\ 
Observation noise: $\bm{\upxi}$ & Normal: $\mathcal{N}(\mathbf{0}, 5 \cdot \mathbf{I}_{50})$ & Sec. \ref{seq4.1} \\
\bottomrule
\multicolumn{3}{l}{\small Note: Cost function parameters investigated parametrically in Section \ref{seq4.3}.} \\
\end{tabular}
\end{table}

The Bayesian inference problem corresponding to monitoring strategy $(z_0)$ is described by the directed graphical model (DGM) in Figure \ref{Fig1} (b). The objective is to infer the deterioration model parameters $\bm{\uptheta}$ directly from thickness loss observations, together with the prediction error standard deviation $\sigma$. The same priors as in Eq. (\ref{eq21}) are used for the model parameters, while $\sigma \sim \text{Half-Normal}(1.0)$. This choice ensures that the standard deviation remains strictly positive while remaining relatively uninformative.

The observation model defined in Eq. (\ref{eq22}) is used to generate 50 thickness loss observations at $t_{\text{insp}} = 15 \ \text{years}$ for each of the 1000 deterioration realisations drawn from the prior. For each prior realisation, NUTS was run with 2000 warm-up steps—to tune the algorithm's hyperparameters and ensure convergence to the invariant distribution—followed by an equal number of draws from the target posterior.
Multiple chains with different seeds were executed for a subset of prior realisations to assess convergence, which was evaluated using the rank-normalised $\hat{R}$ diagnostic. The values of $\hat{R}$ remained consistently below the accepted threshold of 1.01 \citep{Vehtari2021}. The same procedure was followed for all other monitoring strategies, with similar convergence behaviour observed.

\begin{figure}[!t]
	\centering
	\includegraphics[scale=0.9]{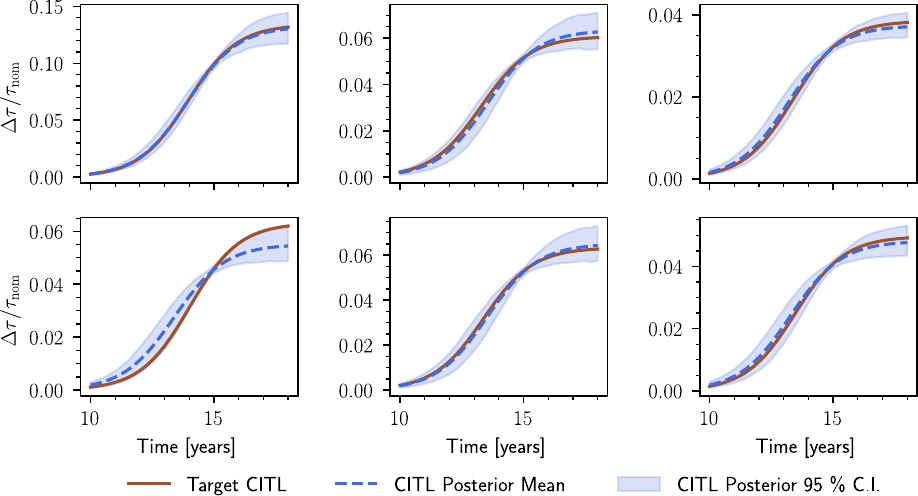}
	\caption{Corrosion-induced thickness loss (CITL) posteriors for six representative, randomly selected thickness loss realisations using monitoring strategy $(z_0)$.}
	\label{Fig6}
\end{figure}

Figure \ref{Fig6} shows the posterior mean and 95\% credible interval of the inferred thickness loss plotted against the corresponding target values—obtained from the prior process—for six indicative, randomly selected realisations of the thickness loss time history. The posterior thickness loss trajectories were obtained by propagating samples from the posterior distribution of the deterioration model parameters through the model defined in Eq.(\ref{eq20}).

The inferred posterior captures the target deterioration curves with good accuracy, while uncertainty is markedly reduced around the inspection time, $t_{\text{insp}}$, as indicated by the narrow credible intervals. It should be noted, however, that the inspection time is critical for inference quality, with earlier inspections expected to yield less accurate results.

The Bayesian inference problem associated with monitoring strategy $(z_1)$ is represented by the DGM in Figure \ref{Fig1} (c). In this case, the quantity of interest is the thickness loss itself, inferred independently at successive points in time. Independent, identically distributed uniform priors are assigned to each time step, namely $\Delta \tau_{k} \sim \mathcal{U}(0.0, 2.0) \ \text{mm}$. This choice effectively bounds the inferred thickness loss within the range of the prior process realisations while preventing the observation model (surrogate) from being queried outside its training domain.

Strain observations were generated for each prior realisation at three-month intervals, with 50 samples per time point, using the linear regression-based surrogate model. Although this constitutes a so-called inverse crime, it is unavoidable due to the nature of the pre-posterior decision analysis. The generated observations were further corrupted with additive Gaussian noise $\bm{\upxi} \sim \mathcal{N}(\mathbf{0}, 5^{2} \cdot \mathbf{I}_{50}) \ \upmu\upepsilon$, where the standard deviation of 5 $\upmu\upepsilon$ reflects the resolution of commonly used strain sensors \citep{Chadha2021}.

The choice of data acquisition frequency was motivated by the slow evolution of thickness loss and by practical considerations related to the observation model. To ensure the validity of the static strain-based model, the vessel must either be in port or sailing in calm seas. These conditions can be reasonably expected to occur within the selected three-month data acquisition interval.

\begin{figure}[!htp]
	\centering
	\includegraphics[scale=0.9]{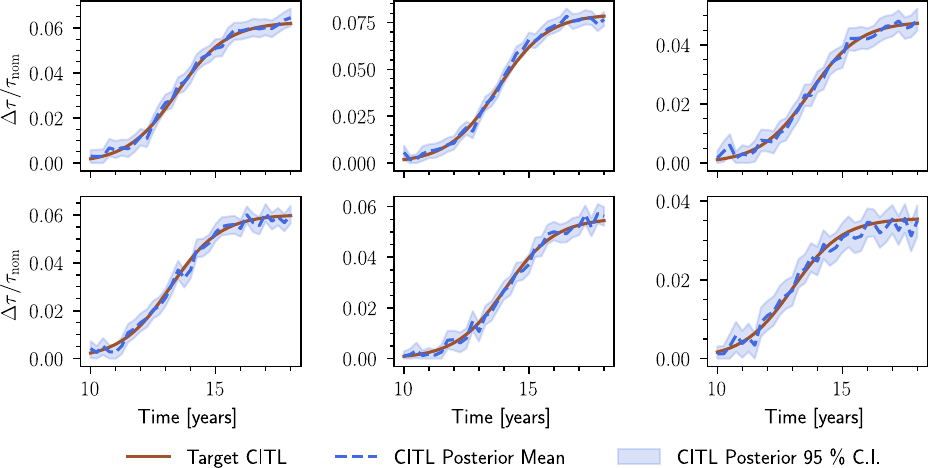}
	\caption{Corrosion-induced thickness loss (CITL) posteriors for six representative, randomly selected thickness loss realisations using monitoring strategy $(z_1)$.}
	\label{Fig7}
\end{figure}

The same hyperparameters were used for the No-U-Turn Sampler as in strategy $(z_0)$. Figure \ref{Fig7} compares the posterior mean and 95\% credible interval of the inferred thickness loss with the corresponding target values for representative deterioration realisations. The observed non-smoothness of the posterior trajectories is expected, as no functional form was imposed on the deterioration model.
Inference performance can again be considered satisfactory, although this approach inevitably introduces minor physical inconsistencies, since the inferred thickness loss is not constrained to evolve monotonically. The posterior mean shown in the upper-right panel of Figure \ref{Fig7} illustrates this typical behaviour.

Monitoring strategies~$(z_2)$ and~$(z_3)$ correspond to the same Bayesian inference problem, represented by the DGM in Figure~\ref{Fig1}~(d). The deterioration model parameter priors follow Eq.~(\ref{eq21}), while the prediction error standard deviation is assigned a $\text{Half-Normal}(1.0)$ prior. The two strategies differ in the number of strain observations available at each data acquisition step, which occurs at three-month intervals. Specifically, a single observation is used at each time step for~$(z_2)$, whereas 50 observations are used for~$(z_3)$. The No-U-Turn Sampler was implemented with the same hyperparameters as in the previous cases.

\begin{figure}[!b]
	\centering
	\includegraphics[scale=0.9]{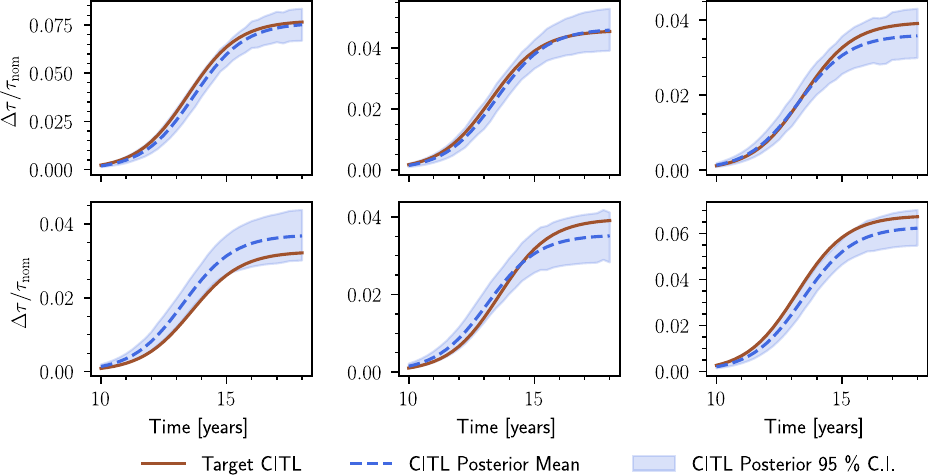}
	\caption{Corrosion-induced thickness loss (CITL) posteriors for six representative, randomly selected thickness loss realisations using monitoring strategy $(z_2)$.}
	\label{Fig8}
\end{figure}
\begin{figure}[!t]
	\centering
	\includegraphics[scale=0.9]{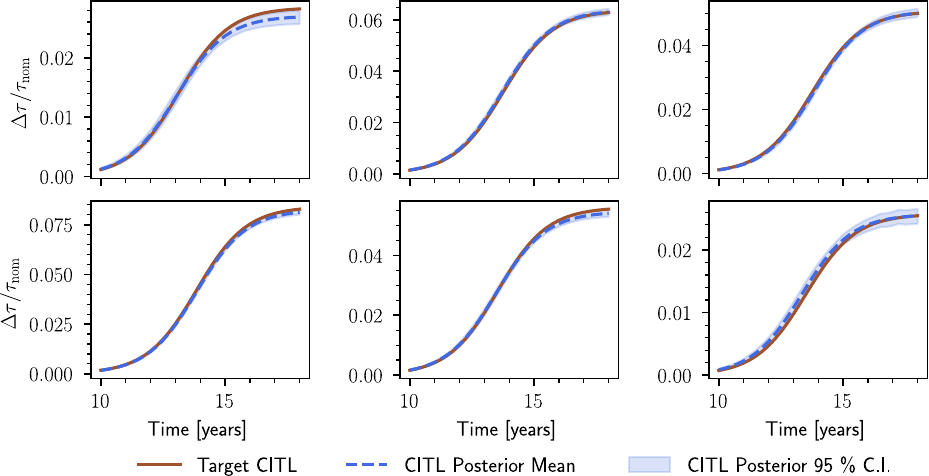}
	\caption{Corrosion-induced thickness loss (CITL) posteriors for six representative, randomly selected thickness loss realisations using monitoring strategy $(z_3)$.}
	\label{Fig9}
\end{figure}

Inference results for representative prior realisations are shown in Figure~\ref{Fig8} for strategy~$(z_2)$ and Figure~\ref{Fig9} for~$(z_3)$. Increasing the number of available observations has a clear positive effect: the posterior for~$(z_3)$ exhibits minimal uncertainty and closely matches the target thickness loss. Conversely, in the sparse-data regime of~$(z_2)$, the posterior variance is higher, though the mean trajectory still approximates the target deterioration curve with reasonable accuracy. Notably, running the No-U-Turn Sampler for~$(z_2)$ required roughly one-fifth of the computational cost of~$(z_3)$ in terms of wall time.

\subsection{Examining the limit state function}\label{seq4.2}

At this stage, it is useful to examine the characteristics of the limit state function and the associated exceedance probability used to define the consequence cost functions. The computational cost of the pre-posterior decision analysis imposes constraints on the number of samples drawn from the thickness loss prior process and from the posterior distributions. The number of posterior samples, $n_{\text{post}}$, primarily affects the robustness of the estimator in Eq. (\ref{eq16}), while the number of prior samples, $n_{\text{prior}}$, influences the Monte Carlo estimators of the prior and pre-posterior consequence costs (Eqs. (\ref{eq6}) \& (\ref{eq10})). In this work, it has been assumed that $n_{\text{prior}} = 1000$ and $n_{\text{post}} = 2000$.

\begin{figure}[!htp]
	\centering
	\includegraphics[scale=1.0]{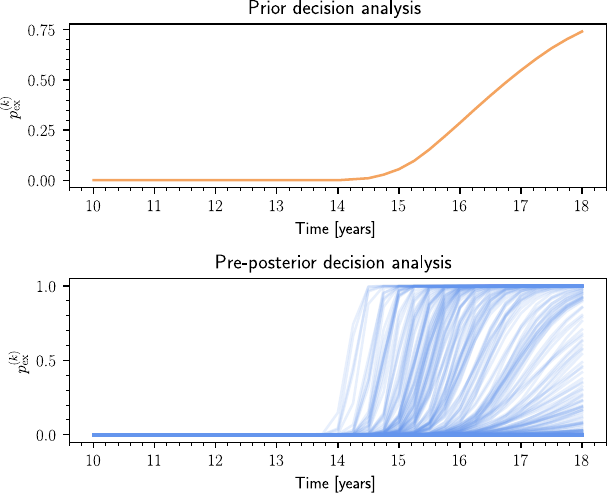}
	\caption{Cumulative exceedance probability over the monitoring period for $\Delta\tau^{(\text{th})} \sim \mathcal{N}\left(1.2, (0.05 \cdot 1.2)^2\right)$ and monitoring strategy $(z_3)$.}
	\label{Fig10}
\end{figure}

Figure \ref{Fig10} illustrates the cumulative exceedance probability, $p^{(k)}_{\text{ex}}$, over the monitoring period, computed using Eq. (\ref{eq17}) for both the prior and the pre-posterior decision analyses. A threshold thickness loss level $\Delta\tau^{(\text{th})} \sim \mathcal{N}\left(1.2, (0.05 \cdot 1.2)^2\right)$ mm is assumed. A Gaussian distribution is selected since it is symmetric and commonly used to describe statistical discrepancy, thereby avoiding unwanted bias in modelling epistemic uncertainty---implicitly achieved through the threshold distribution. The 5\% coefficient of variation ensures that threshold realisations remain close to the prescribed mean value---the threshold level---and prevents generating physically implausible thresholds outside the prior range.

As expected, for the prior decision analysis, $p^{(k)}_{\text{ex}}$, reduces to a single monotonically increasing curve that becomes non-zero after approximately the 14$^{\text{th}}$ year of vessel operation. At this point, thickness loss levels approach the maintenance-related threshold. A similar trend can be seen in the cumulative exceedance probability estimates from the pre-posterior analysis corresponding to monitoring strategy $(z_3)$, which are representative of all cases.

\begin{figure}[!t]
\centering
\includegraphics[scale=0.82]{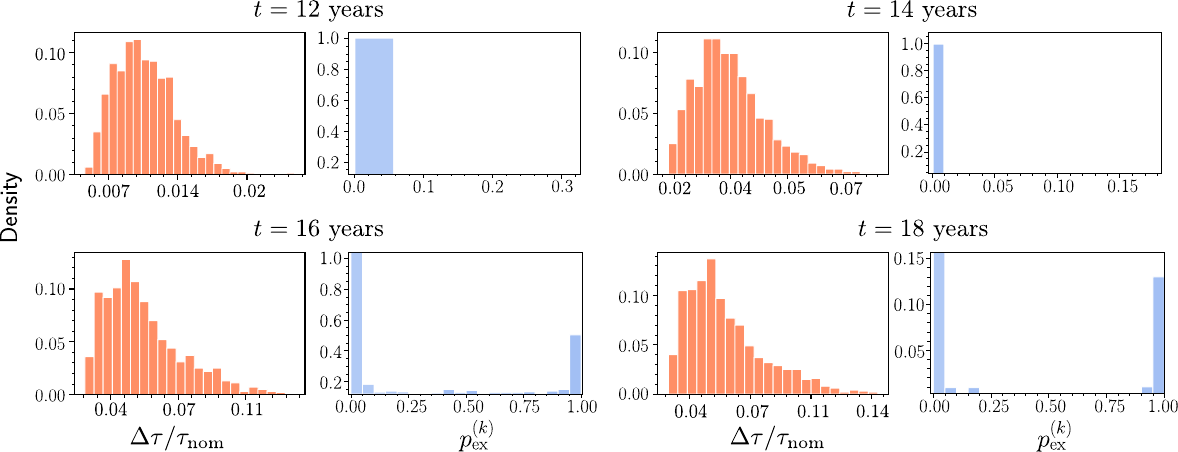}
\caption{Thickness loss prior realisation histograms (orange) over the monitoring period and corresponding cumulative exceedance probability histograms (blue) from the pre-posterior decision analysis for monitoring strategy $(z_3)$.}
\label{Fig11}
\end{figure}

A key feature of the $p^{(k)}_{\text{ex}}$ curves in the pre-posterior case is that, for a large proportion of deterioration realisations, the exceedance probability transitions rapidly from near-zero to one. This reflects the narrow thickness loss posteriors shown in Figure \ref{Fig9}. Consequently, the exceedance probability estimator oscillates between 0 and 1, depending on the time and the decision threshold. Figure \ref{Fig11} illustrates this behaviour by comparing histograms of prior thickness loss realisations at selected time points with the corresponding $p^{(k)}_{\text{ex}}$ histograms obtained from the pre-posterior decision analysis.

At times when thickness loss levels have not yet reached the threshold, as shown in the upper panel of Figure \ref{Fig11}, the estimator effectively tends towards zero. As time progresses, the lower panel of Figure \ref{Fig11} shows that a proportion of deterioration scenarios still remain below the threshold. However, only a few cases yield intermediate values of $p^{(k)}_{\text{ex}}$, while in most scenarios it approaches 1. This indicates that the estimator of Eq. (\ref{eq16}) behaves like a discrete variable due to the low posterior uncertainty and the monotonically increasing nature of the deterioration model. From a practical standpoint, it also confirms that the adopted number of posterior samples is sufficient to ensure the reliability of the $p^{(k)}_{\text{ex}}$ estimator.

The high quality of the inference results suggests that the posterior thickness loss process closely matches the prior realisations. Having established the robustness of the estimator for the chosen number of posterior samples, attention now turns to assessing whether the number of prior samples is also sufficient. An empirical approach was adopted for this purpose, tailored to the characteristics of the present analysis.

Given the low computational cost of generating prior realisations, $10^6$ thickness loss realisations were drawn for the monitoring period. Histograms of these realisations at different points in time were used to visually identify an appropriate probability distribution. Using this distribution, interval exceedance probabilities were then estimated for multiple threshold values and time instances, again using $10^6$ samples to ensure a high level of confidence.

\begin{figure}[!b]
	\centering
	\includegraphics[scale=0.9]{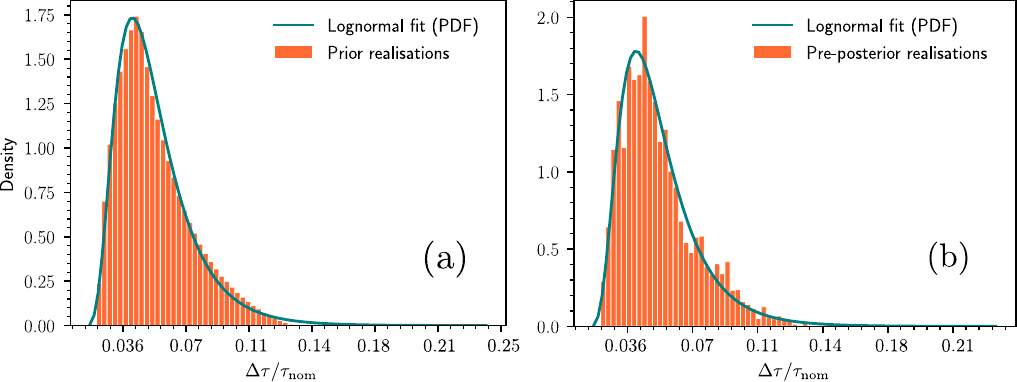}
	\caption{Histograms and fitted log-normal probability distributions of thickness loss for (a) prior process realisations and (b) posterior realisations for the entire pre-posterior analysis at $t = 16 \ \text{years}$.}
	\label{Fig12}
\end{figure}

Figure \ref{Fig12} (a) shows a representative histogram obtained using this procedure for $t = 16 \ \text{years}$ alongside the corresponding log-normal distribution fit. Figure \ref{Fig12} (b) presents a histogram generated by combining all posterior samples obtained during the pre-posterior analysis at the same time point. These posterior realisations correspond to $n_{\text{prior}} = 1000$ deterioration scenarios. Their low individual variance (see Fig \ref{Fig9}) indicates that they effectively act as replicas of the corresponding prior realisation, resulting in an exceedance probability estimator analogous to a bootstrap estimator. Nevertheless, a log-normal probability distribution was also fitted to the posterior realisations, and exceedance probabilities were estimated for different time points and threshold values.

For a threshold level  $\Delta\tau^{(\text{th})} \sim \mathcal{N}\left(1.5, (0.05 \cdot 1.5)^2\right)$ mm at 16 years, the exceedance probability estimated using the prior was approx. 2.21\%, while that based on the posterior samples approx. 1.94\%. This difference reflects the divergence between estimators in the upper tail of the distributions, where discrepancies are most pronounced. At lower threshold levels however, the estimators are virtually identical. Given the consequence cost formulation (see Figure \ref{Fig2}), exceedance probabilities of this magnitude are not expected to influence maintenance decisions, as they fall below the $\hat{\text{P}}^{\text{(c)}}_{\text{ex,th}}$ levels of interest. It can therefore be concluded that the assumed sample sizes are sufficient for the purposes of this study.

\subsection{Investigating the effect of the decision threshold}\label{seq4.3}

We begin the investigation of the Value of Information from thickness loss monitoring by examining how the decision threshold influences the expected reward-to-investment risk ratio, $\lambda$, across the different strain-based monitoring strategies. To do so, both intrinsic and extrinsic costs must first be defined. Owing to the lack of specific cost data, the intrinsic cost—which encompasses installation and operation and maintenance expenses—is specified heuristically.

Monitoring strategy $(z_2)$ is assumed to have the lowest intrinsic cost and is therefore used as the reference. This reflects its lower data requirements compared with the other two strain-based monitoring strategies. Accordingly, we set $\mathcal{C}(z_2) = 0.1\cdot\mathcal{C}_{\text{max}}$ and $\mathcal{C}_{\text{O\&M}}(z_2) = 0.01\cdot\mathcal{C}(z_2)$, where $\mathcal{C}_{\text{max}}$ denotes the overall maximum consequence cost, equal to 1. These relative cost assignments are based on engineering judgement and on the figures reported by Chadha et al. \citep{Chadha2021}.

The higher data requirements of strategies $(z_1)$ and $(z_3)$—each using 50 strain observations per data acquisition time instead of one—are assumed to increase installation costs by 10\%, i.e., $\mathcal{C}(z_1) = \mathcal{C}(z_3) = 1.1 \cdot \mathcal{C}(z_2)$. O\&M costs are assumed to scale proportionally with the relative increase in computation cost, yielding $\mathcal{C}_{\text{O\&M}}(z_1) = 2 \cdot \mathcal{C}_{\text{O\&M}}(z_2)$ and $\mathcal{C}_{\text{O\&M}}(z_3) = 5 \cdot \mathcal{C}_{\text{O\&M}}(z_2)$. Finally, a constant yearly inflation rate of $r = 2\%$ is assumed.

For the extrinsic costs, two consequence cost definitions are considered for the decision to repair, denoted as $\mathcal{R}_{1}\left(d_1\right) \ \text{and} \ \mathcal{R}_{2}\left(d_1\right)$. These are defined for $\left( \min\mathcal{R}_{1}\left(d_1\right), \hat{\text{P}}^{\text{(c)}}_{\text{ex,th}} \right) = \left( 0.33, 0.2 \right)$ and $\left( \min\mathcal{R}_{2}\left(d_1\right), \hat{\text{P}}^{\text{(c)}}_{\text{ex,th}} \right) = \left( 0.15, 0.1 \right)$. The first, $\mathcal{R}_{1}\left(d_1\right)$, can be interpreted as representing a more risk-seeking decision-maker who assigns greater cost to false calls (i.e., unnecessary repairs) and is willing to accept a higher maintenance-related threshold. Conversely, $\mathcal{R}_{2}\left(d_1\right)$ can be interpreted as reflecting a more risk-averse attitude.

\begin{figure}[!htb]
	\centering
	\includegraphics[scale=0.82]{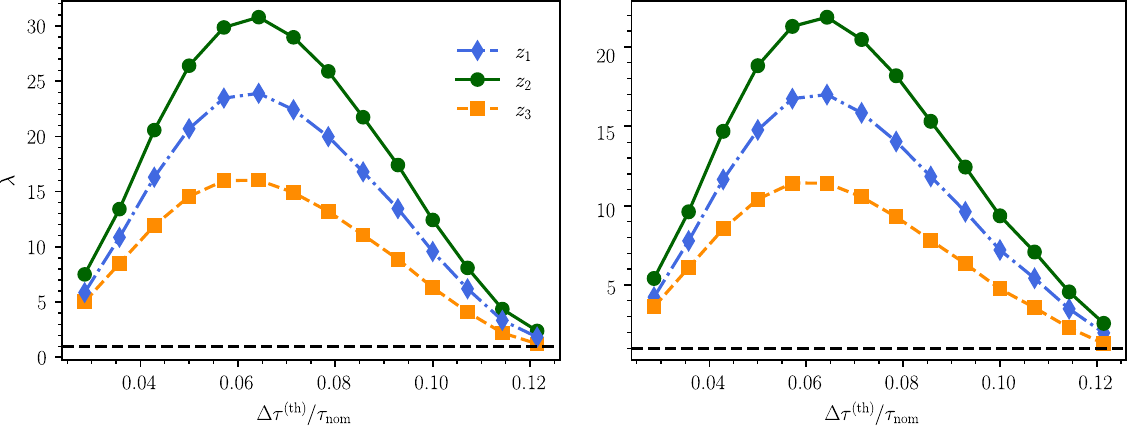}
	\caption{Expected reward-to-investment risk ratio as a function of the decision threshold mean value for different strain-based monitoring strategies. Left panel corresponds to $\mathcal{R}_{1}\left(d_1\right)$ and right panel to $\mathcal{R}_{2}\left(d_1\right)$.}
	\label{Fig13}
\end{figure}

Figure \ref{Fig13} presents the expected reward-to-investment risk ratio for both consequence cost definitions across different thickness loss threshold levels. The three monitoring strategies exhibit similar trends, with investment in strain-based monitoring being optimal at intermediate threshold levels. At lower thresholds, the value of acquiring monitoring information diminishes, as the prior process exhibits lower variance and more clustered realisations (see Figure  \ref{Fig3}). Conversely, at higher thresholds the benefit of investing in monitoring becomes marginal, becasue $p^{(k)}_{\text{ex}} \rightarrow 1$ for a large proportion of deterioration scenarios, leading to substantially increased compounded repair costs.

Among the monitoring strategies, $(z_2)$, i.e., the reference case, emerges as the most favourable. This result stems from the high accuracy and low uncertainty of the corresponding thickness loss posteriors, which leave the initial cost assignments as the primary differentiating factor. A more thorough investigation into cost definition, ideally using real-world data from operating vessels or other sources, would be required to draw more nuanced conclusions on this front. Such an analysis lies beyond the scope of the present study.

Comparison of the two consequence cost definitions confirms the intended interpretation of the decision-maker's risk profile: the risk-seeking case $\left( \mathcal{R}_1 \left( d_1 \right) \right)$ shows potential to achieve greater maximum rewards. Importantly, however, the marginal behaviour of the expected reward-to-investment risk ratio remains similar regardless of risk attitude, underscoring the general benefit of acquiring monitoring information.

\begin{figure}[!t]
	\centering
	\includegraphics[scale=0.9]{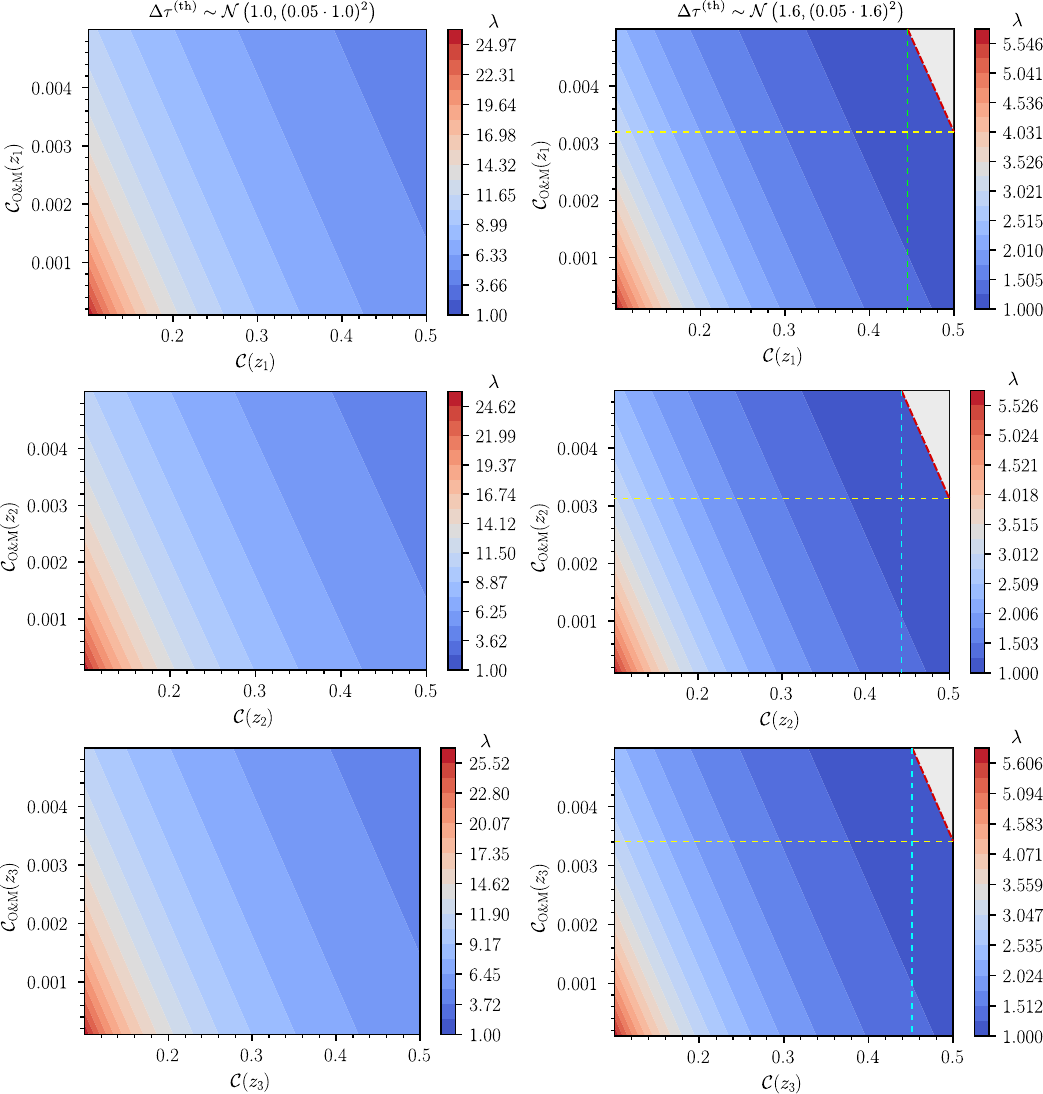}
	\caption{Expected reward-to-investment risk ratio maps for different intrinsic cost assignments.}
	\label{Fig14}
\end{figure}

\subsection{Investigating cost margins justifying monitoring}\label{seq4.4}

To further explore the effect of intrinsic costs on the potential value of strain-based thickness loss monitoring, the expected reward-to-investment risk ratio, $\lambda$, was computed across a broader range of $\mathcal{C}(z_i)$ and $\mathcal{C}_{\text{O\&M}}(z_i)$ values. Specifically, $\mathcal{C}(z_i)$ was varied within $\left[ 0.1 \cdot \mathcal{C}_{\max}, 0.5 \cdot \mathcal{C}_{\max} \right]$, while $\mathcal{C}_{\text{O\&M}}(z_i)$ was set to $0.01\cdot\mathcal{C}(z_i)$. The ratio $\lambda$ was then evaluated over a $100 \times 100$ uniform grid made up of combinations of these two costs.

Results for the intermediate threshold level confirm that strain-based thickness loss monitoring can provide benefits to maintenance planning for ship hull structures for a wide range of intrinsic cost assignments. As the threshold increases, however, a non-feasible cost region appears where $\lambda < 1$, indicated by grey shading in Figure \ref{Fig14}. Notably, the feasible and non-feasible intrinsic cost regions are linearly separable, a direct consequence of the definition of linear consequence cost functions. This is a useful property that enhances interpretability, allowing stakeholders to derive straightforward decision rules.

Under identical intrinsic cost assumptions, the different monitoring strategies contribute nearly identical value to thickness loss monitoring, reinforcing the earlier conclusion that their relative benefit is primarily governed by the cost structure. Figure \ref{Fig14} illustrates these results for two threshold levels---one representing intermediate deterioration and the other a more severe case---using the risk-averse definition of the decision-to-repair consequence cost function, $\mathcal{R}_2\left(d_1\right)$.

\subsection{Investigating the utility of inspections} \label{seq4.5}

At this stage, strain-based thickness loss monitoring has been shown to be a reasonable investment to support maintenance planning decisions for ship hull structures. However, to comprehensively assess its value, it must also be compared with the Value of Information (VoI) associated with typical inspection strategies. To this end, the relative risk-adjusted reward metric, $\chi \left( z_0, z_i \right)$, defined in Eq. (\ref{eq15}), is employed, where $z_i$ denotes the strain-based monitoring strategies and $z_0$ represents the inspection-only strategy discussed earlier.

To calculate the expected reward-to-investment risk ratio for inspections, minor adjustments are required to the definition of the expected lifecycle pre-posterior loss in Eq. (\ref{eq11}). Since the scenario involves a single inspection event, it is assigned a dedicated intrinsic cost, $\mathcal{C}_{\text{insp}}$, while the summation defining the extrinsic cost in Eq. (\ref{eq11}) reduces to a single term, as defined in Eq. (\ref{eq10}). With this modification, the implementation described in Figure \ref{algo_flow} can be followed directly to estimate the VoI for inspection-only data.

\begin{figure}[!htb]
	\centering
	\includegraphics[scale=0.8]{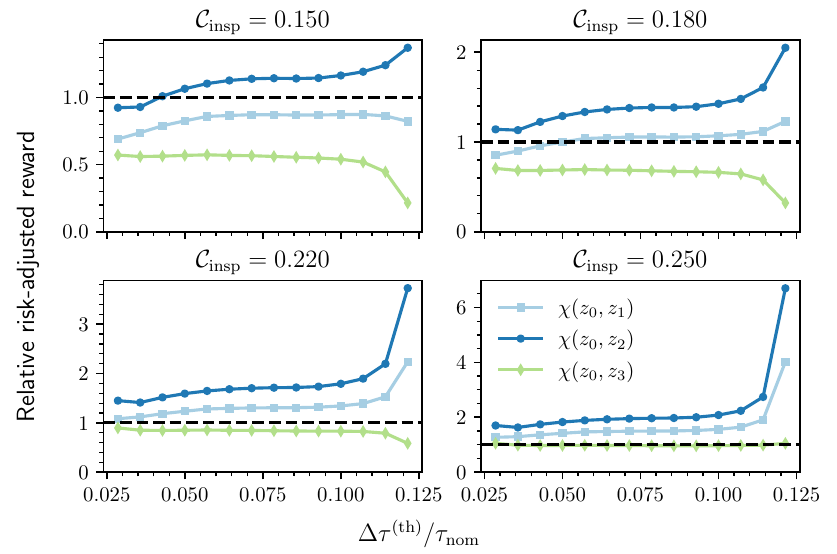}
	\caption{Relative risk-adjusted reward between strain-based monitoring strategies and inspection-only monitoring assuming consequence cost function $\mathcal{R}_{1}\left(d_1\right)$.}
	\label{Fig15}
\end{figure}

\begin{figure}[!t]
	\centering
	\includegraphics[scale=0.8]{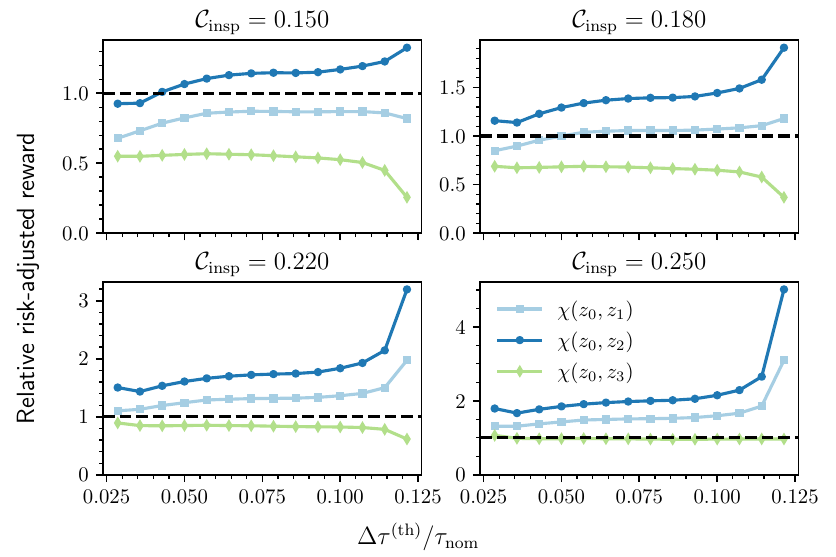}
	\caption{Relative risk-adjusted reward between strain-based monitoring strategies and inspection-only monitoring assuming consequence cost function $\mathcal{R}_{2}\left(d_1\right)$.}
	\label{Fig16}
\end{figure}

The results of this comparison are shown in Figures \ref{Fig15}--\ref{Fig16}, which plot the relative risk-adjusted reward between the strain-based monitoring strategies and the inspection-only scenario for different maintenance-related thickness loss thresholds. The risk-seeking profile, defined by the decision-to-repair consequence cost $\mathcal{R}_1 (d_1)$, was used to obtain Figure \ref{Fig15}, while $\mathcal{R}_2 (d_1)$—representing the more risk-averse decision-maker—was used for Figure \ref{Fig16}. For the strain-based monitoring strategies, the relative cost assignments discussed in Section \ref{seq4.3} have been assumed here as well.


The black dashed line at $\chi \left( z_i, z_0 \right) = 1$ marks the boundary above which investment in strain-based monitoring is more beneficial than a conventional inspection-only strategy. It should be emphasised that this finding---and all conclusions here---hold only under the assumed inspection strategy and monitoring period. Across both consequence cost function definitions and all considered threshold levels and inspection costs, $\chi \left( z_i, z_0 \right) > 0$. This indicates that, as with strain-based monitoring (see Figure \ref{Fig13}), investment in an inspection-only monitoring strategy can also be considered beneficial. However, this comes with an important caveat: the inspection time. Results would differ if inspections occurred earlier in the deterioration process—likely yielding less favourable outcomes.

The intrinsic cost of strain-based monitoring and the inspection cost are the main factors determining which strategy is preferable. Under the assumed intrinsic cost assignments, monitoring strategy $(z_2)$ appears optimal for most scenarios. Its relative value increases with higher threshold levels, underscoring the advantage of a monitoring system capable of tracking deterioration more frequently. Regarding risk-perception, a more risk-averse decision-maker tends to assign lower utility to strain-based monitoring. However, this utility increases as inspection costs rise, regardless of risk profile.

Finally, it should be acknowledged that inspection costs are inherently volatile, being influenced by numerous external factors such as global trade conditions and shipping rates. Accounting for these factors—individually or collectively—is highly challenging and beyond the present scope, but constitutes a promising direction for future research.

\section{Concluding remarks} \label{seq5}

This work presented the first comprehensive Value of Information-based assessment of strain-based monitoring for corrosion-induced thickness loss in ship hull structures. The proposed Bayesian pre-posterior decision analysis framework employed consequence cost functions defined in terms of exceedance probabilities relative to maintenance-related thickness loss thresholds, thereby enabling implicit modelling of decision-maker risk perception.

Applying the framework to a high-fidelity numerical model of a commercial vessel, the following key findings were demonstrated:
\begin{itemize}
    \item Strain-based monitoring of thickness loss can provide quantifiable value to maintenance decision-making across a range of deterioration scenarios and cost assumptions, with an optimal expected reward-to-investment risk ratio at intermediate threshold levels (Figures \ref{Fig13} \& \ref{Fig14}).
    \item The optimal monitoring strategy depends critically on the interplay between system intrinsic costs, consequence cost definitions, and maintenance-related threshold levels.
    \item Under the assumptions of this analysis, continuous strain monitoring with limited data acquisition yields better risk-adjusted returns than either more data-intensive strategies or conventional inspection-only approaches (Figures \ref{Fig15} \& \ref{Fig16}).
\end{itemize}

The credibility of the proposed framework stems from its consistent decision-theoretic formulation, which provides a rational and transparent basis for linking probabilistic inference, reliability assessment, and maintenance decision-making.

Despite these contributions, the study is subject to certain limitations. The lack of detailed historical data on inspection and maintenance costs necessitated the use of normalised cost definitions, introducing interpretational caveats concerning the generalisability of the quantitative results. Furthermore, while the binary decision set and linear consequence costs were deemed reasonable for the present purpose, future research should explore more realistic action/decision settings and cost formulations.

The use of synthetic strain data, appropriate for pre-posterior analysis and simultaneously necessitated by the lack of operational monitoring data, implies quantitative Value of Information results are case-specific. Nonetheless, the framework itself is transferable, and relative comparisons between monitoring strategies are expected to remain robust under moderate modelling uncertainty. The framework assumes reliable sensor operation throughout the monitoring period and does not explicitly model sensor degradation or intermittent failures, which could affect inference quality and decision outcomes.
	
Incorporating sensor reliability within a pre-posterior optimal sensor placement framework (similar to \citep{Yang2023_MSSP}) represents an important step towards real-world deployment. Future work integrating operational data from instrumented vessels would also strengthen confidence in quantitative predictions. Additionally, explicit modelling of repair actions and their outcomes constitutes a key direction for future research. Such advances could further establish the Value of Information as a practical metric for optimising maintenance and inspection scheduling across engineering structures.

\section*{Declaration of conflicting interests}
The authors declared no potential conflicts of interest with respect to the research, authorship, and/or publication of this article.

\section*{Funding}
The authors received no financial support for the research, authorship, and/or publication of this article.


\clearpage
\appendix
\section*{Appendix A. Nomenclature}
\addcontentsline{toc}{section}{Appendix A. Nomenclature}
\begin{longtable}{@{}p{3.1cm}p{12.1cm}@{}}
\toprule
\textbf{Symbol} & \textbf{Description} \\
\midrule
\endfirsthead
\toprule
\textbf{Symbol} & \textbf{Description} \\
\midrule
\endhead
\bottomrule
\endfoot

$\Delta \mathcal{T}$ & Corrosion-induced thickness loss (CITL) random variable. \\
$\Omega_{\Delta \mathcal{T}}$ & Sample space of the random variable $\Delta \mathcal{T}$. \\
$\Delta \tau$ & Realisation of the thickness loss random variable $\Delta \mathcal{T}$. \\
$t_k$ & Discrete time point. \\
$\mathcal{F}(t; \bm{\uptheta})$ & Deterioration model describing temporal evolution of thickness loss. \\
$\bm{\uptheta}$ & Vector of deterioration model parameters. \\
$d_{\theta}$ & Dimension of parameter vector $\bm{\uptheta}$. \\
$\mathcal{H}(\cdot)$ & Observation (measurement) model mapping structural states to measurable quantities. \\
$\mathbf{y}_k$ & Observation vector at discrete time $t_k$. \\
$\bm{\upxi}$ & Observation noise random vector (aleatoric uncertainty). \\
$\psi$ & Process noise random variable (epistemic uncertainty in the state model). \\
$p(\cdot)$ & Probability density function. \\
$\mathcal{R}(d_j, \bm{\uptheta})$ & Consequence cost function for decision $d_j$ given state $\bm{\uptheta}$. \\
$d_j$ & Candidate maintenance decision ($d_0$: no repair, $d_1$: repair). \\
$\ind \left[ \cdot \right]$ & Indicator function. \\
$p_{\mathrm{ex}}^{(k)} = \hat{\text{P}}^{\text{(c)}}_{\text{ex}}(t_k)$ & Cumulative exceedance probability of maintenance-related thickness loss threshold at time $t_k$. \\
$\hat{\text{P}}^{\text{(i)}}_{\text{ex}} (t_k)$ & Interval exceedance probability estimator at time $t_k$. \\
$\Delta \mathcal{T}^{\text{(th)}}$ & Random variable representing the maintenance threshold thickness loss. \\
$\min \mathcal{R}(d_1)$ & Baseline (false-alarm) cost parameter for the repair decision. \\
$\hat{\text{P}}^{(c)}_{\mathrm{ex,th}}$ & Risk-perception threshold parameter used to define cost functions. \\
$z_i$ & Candidate information/monitoring strategy (e.g., inspection-only, strain-based). \\
$\mathbf{y}_{z_i}$ & Observation vector for candidate monitoring strategy $z_{i}$. \\
$r_k$ & Annual inflation rate at time $t_k$. \\
$\mathcal{C}^{(k)}_{\mathrm{prior}}$ & Expected loss (Bayes risk) at time $t_{k}$ under prior decision analysis. \\
$\mathcal{C}^{(k)}_{\mathrm{pre}}(z_i)$ & Expected loss at time $t_{k}$ under pre-posterior analysis with strategy $z_i$. \\
$\mathcal{C}_{\mathrm{prior}}$ & Lifecycle expected loss under prior decision analysis. \\
$\mathcal{C}_{\mathrm{pre}}(z_i)$ & Lifecycle expected loss under pre-posterior analysis with strategy $z_i$. \\
$\mathcal{C}_{S}(z_i)$ & Expected cost savings due to monitoring strategy $z_i$. \\
$\mathcal{C}(z_i)$ & Installation (intrinsic) cost of strategy $z_i$. \\
$\mathcal{C}_{\mathrm{O\&M}}(z_i)$ & Operation \& maintenance cost of strategy $z_i$. \\
$\mathrm{EVOI}(z_i)$ & Expected value of information for strategy $z_i$. \\
$\lambda(z_i)$ & Expected reward-to-investment risk ratio for strategy $z_i$. \\
$\chi(z_1,z_2)$ & Relative risk-adjusted reward comparing strategies $z_1$ and $z_2$. \\
$k,\,i,\,j,\,n,\,m$ & Indices for time steps, estimators, decisions, Monte Carlo samples, and years. \\
$K$ & Number of decision times considered in lifecycle analysis. \\
$M$ & Number of operation years considered for O\&M costs. \\
$p(\bm{\uptheta})$ & Prior pdf for deterioration-model parameters. \\
$p(\bm{\uptheta}\mid \mathbf{y}_{z_i})$ & Posterior pdf given observations under strategy $z_i$. \\
$n_{\mathrm{prior}},\,n_{\mathrm{post}}$ & Monte Carlo sample sizes for prior and posterior stages. \\
$\hat{R}$ & Rank-normalised potential scale reduction factor: convergence diagnostic for Markov Chain Monte Carlo. \\
$\mathbf{y}_{\text{insp}}$ & Observation vector for inspection-only monitoring strategy. \\
$t_{\text{insp}}$ & Time of inspection. \\
$\alpha, \beta, \gamma$ & Logistic-type deterioration model parameters. \\
$\sigma$ & Standard deviation of the zero-mean, uncorrelated Gaussian prediction error model. \\
$\varepsilon_{xx}$ & Longitudinal strain at a monitored location. \\

\end{longtable}

\clearpage
\bibliographystyle{elsarticle-num-names}
\bibliography{VoI_Paper_Refs.bib}

\end{document}